\documentclass[10pt,a4paper]{article}
\usepackage[utf8]{inputenc}

\title{Synchronized}
\author{gyuszko2000 }
\date{August 2020}

\usepackage{amsmath,amssymb,graphicx}

\usepackage{latexsym,amssymb,graphics,graphicx,epsfig,color,enumerate}
\usepackage{xspace}
\usepackage{amsthm}

\usepackage{caption}
\usepackage{subcaption}

\usepackage{t1enc}

\newtheorem{thm}{Theorem }

\newtheorem{claim}{Claim }
\newtheorem{defn}{Definition }
\newtheorem{problem}{Problem }

\newtheorem{cor}{Corollary }
\newtheorem{conj}{Conjecture }

\title{Synchronized Traveling Salesman Problem 
}

\author{
Gyula Pap
\thanks{
The first author was supported in part by the CELSA Research Fund, and . Member of MTA-ELTE Egerv\'ary Research Group, Department of Operations Research, E\"otv\"os University, P\'azm\'any P\'eter s\'et\'any 1/C, Budapest, Hungary, H-1117. E-mail: {\tt gyuszko@cs.elte.hu}. This research is supported by the Hungarian National Research, Development and Innovation Office grant NKFI-132524. 
}
\and 
J\'ozsef Varny\'u
\thanks{The project was supported by the European Union, co-financed by the European
Social Fund (EFOP-3.6.3-VEKOP-16-2017-00002). }
}

\date{September 2018, revised October 2020}

\begin{document}

\maketitle

\begin{abstract}
We consider a variation of the well-known traveling salesman problem in which there are multiple agents who all have to tour the whole set of nodes of the same graph, while obeying node- and edge-capacity constraints require that agents must not "crash". We consider the simplest model in which the input is an undirected graph with all capacities equal to one. A solution to the synchronized traveling salesman problem is called an "agency". Our model puts the synchronized traveling salesman problem in a similar relation with the traveling salesman problem as the so-called evacuation problem, or the well-known dynamic flow (flow-over-time) problem is in relation with the minimum cost flow problem. 

We measure the strength of an agency in terms of number of agents which should be as large as possible, and the time horizon which should be as small as possible. Beside some elementary discussion of the notions introduced, we establish several upper and lower bounds for the strength of an agency under the assumption that the input graph is a tree, or a 3-connected 3-regular graph. 
\end{abstract}

\section{Introduction}

The purpose of this paper is to explore and study some new problems that are defined from a natural merger of the traveling salesman problem, and the dynamic flow problem. 

Our first starting point is the traveling salesman problem, which is one of the best known problems from combinatorial optimization, and there has been a wide range of variations to the original traveling salesman problem that have been investigated over the decades. In this paper we would consider a variation of the traveling salesman problem that has multiple salesman touring the same network. Of course earlier research has been done with multiple salesmen, including the so-called multiple traveling salesmen problem (Laporte, Norbert, \cite{Laporte1980}), vehicle routing problems (Christofides, Mingozzi, Toth, \cite{Christofides1981}), price-collecting traveling salesman problem (Balas, \cite{Balas1989}), and others. The problem variation considered in this paper will be completely different from those. 

The second starting point is the dynamic flow problem first introduced by Ford, Fulkerson \cite{FF58}, also known as flows over time, which came as a variation to the flow problem in which we consider a flow as a dynamic notion that develops over time. See Skutella, \cite{Skutella2009} and Kotnyek \cite{kotnyek} for a broad survey about this topic.  The idea is to replace the static notion of a flow by a dynamic notion in which flow is described as the time-dependent motion of particles, either discrete, or continuous. The evacuation problem, quickest transshipment problem (see Hoppe, Tardos \cite{hoppetardos}), fastest flow problem, dynamic transportation \cite{bookbinder}, and many others are special cases or variations of the dynamic flow problem. 

In this paper we would develop a notion of dynamic (or synchronized) traveling salesmen, that correspond to the original traveling salesman problem in a similar way as dynamic flows correspond to flows. The idea, basically, is to find a way to put several salesmen, "agents", in a network so that they all need to perform a traveling salesman tour, and the lot of them must obey network capacities. In this setting we may consider two different kind of objectives: try to add as many agents to the networks as possible, or try to find tours for the agents to finish in as short time as possible. Of course the tours of the agents must be coordinated between each other in order to let them obey the network capacities, thus our wording to call this problem the synchronized traveling salesmen problem. One case of this problem has been considered in \cite{komal}. A more precise definition follows below.

\section{Problem Setting}

Let $G=(V,E)$ be an undirected graph with $n=|V|$. A sequence $v(0), v(1),$ $v(2), \cdots ,v(T)$ of nodes $v(t)\in V$ is called a {\bf walk (with parking)} if for all $t=0,1,\cdots ,T-1$ we have either $v(t)=v(t+1)$ or $v(t)v(t+1)\,\in E$. A sequence $v(0), v(1), v(2), \cdots ,v(T)$ of nodes $v(t)\in V$ is called a {\bf walk (without parking)} if for all $t=0,1,\cdots ,T-1$ we have $v(t)v(t+1)\,\in E$. Most of the time in this paper, we will consider walks with parking, so even if not expressly written, a {\bf walk} means a walk with parking. For a walk, $T$ is called the {\bf time horizon}. 

A sequence $v(0), v(1), v(2), \cdots ,v(T)$ of nodes $v(t)\in V$ is called a {\bf traveling salesman tour (with/ without parking)}, if it is a walk (with/without parking) in the graph such that every node appears at least once, and the tour returns to its initial node, that is, $v(0)=v(T)$. A traveling salesman tour with parking is called a {\bf tour}, for short. As a tour returns to its initial node, we may consider a tour by time units modulo $T$ (which is similar to picturing a tour as if it were to repeat all over after the time horizon). 

Of course in the usual setting of the traveling salesman problem, there is no need for parking, because it is just a waste of time or cost; here in our setting, however, parking may be needed to avoid two salesmen of crashing into each other: one of them would wait until the other one passes a node or an edge, and move on afterwards. Vaguely speaking: the point is that we introduce a setting in which there are multiple salesmen touring the same graph at the same time so that they are not allowed to crash into each other. In this setting it makes a lot of sense to allow parking, and this is what we do in this paper. 

In the synchronized traveling salesman problem, we consider an {\bf "agency"} of a number of salesmen that each one of them has to do a tour with the same time horizon, though they need to start from different initial nodes, and must not "crash" into each other. Essentially there is a unit capacity for each node or each edge. More precisely, we define an agency as follows. 

\begin{defn}
Let $k,T\in \mathbb{Z_+}$ be positive integers. $k$ denotes the number of salesmen, or {\bf agents}, and $T$ denotes the joint time horizon. Let $a_i(t)\in V$ be the node where agent $i$ is supposed to be at time $t$, where $i=1,2,\cdots ,k$, and $t=0,1,\cdots ,T$. This is called an {\bf agency} with time horizon $T$ and $k$ agents if for any fixed $i$, $v_i(t)$ is a tour (with parking). In practical terms, each $i$ denotes an agent that moves along the unit-length edges of the graph, so that every agent makes a traveling salesman tour of time horizon $T$. 
\end{defn}

\begin{defn}
If $i$ and $j$ are agents from the same agency, $i\ne j$, then we say that these agents $i$ and $j$ {\bf crash in a node} $v$ at time $t$ if $v=a_i(t)=a_j(t)$. In practical terms, this may be understood as the two agents walking around the graph, and at time $t$ they both occupy in the same node $v$. 
\end{defn}

\begin{defn}
If $i$ and $j$ are agents from the same agency, $i\ne j$, then we say that these agents $i$ and $j$ {\bf crash in an edge} $uv\in E$ at time $t$ if $v=a_i(t)=a_j(t+1), u=a_i(t+1)=a_j(t)$ or $v=a_i(t+1)=a_j(t), u=a_i(t)=a_j(t+1)$. In practical terms, this may be understood as the two agents walking around the graph, thinking of edges as links of a unit length, and at time $t$ they both use the same edge, entering from opposite ends, and thus meeting in (the middle of) the edge. 
\end{defn}

\begin{defn}
An agency is called a {\bf feasible agency} if there is no crash between any pair of agents in neither an edge nor a node. In practical terms, this may be understood as a set of agents moving along the unit-length edges of the graph so that they avoid crashing into each other, but each of them manages to visit every node at least once, before finally arriving at their respective nodes of origin. 
\end{defn}

Given the definition of a feasible agency, we would like to set up a measure the "strength" of an agency. There are two options: we may try to maximize the number of agents, or we may try to minimize the time horizon. Anyway, for any $G,k,T$ there is always the question: is there a feasible agency in graph $G$ with $k$ agents and time horizon $T$? 

\begin{problem}
Given $G,k,T$, decide whether there is a feasible agency in $G$ of $k$ agents under time horizon $T$?
\end{problem}

In Problem 1 if the answer is "yes" for a given $G,k,T$, then -- because parking is allowed -- the answer would be "yes" for any other instance $G,k',T'$ when $k'\le k$ and $T'\ge T$. For a given $k$ we may want to determine the smallest $T$ that admits a feasible agency, and for a given $T$ we may want to determine the largest $k$ that admits a feasible agency. Though these problems are hard to solve in general, we will look for some other ways how to measure the strength of an agency, thus we introduce the following parameters. 

\begin{defn}
For an agency as above, let $\alpha _1:=n/k$, let $\alpha _2:=T/n$. The {\bf strength} of the agency is given by $\alpha :=\max \{\alpha _1,\alpha _2\}$. 
\end{defn}

Notice that for any connected graph there is an agency of $k\ge 1$ and $T\le 2n-2$. Also notice that $k\le n$, and $T\ge n$, for any agency. A more interesting connection is with the Hamiltonian cycle problem. 

\begin{claim}\label{1}
 For a given graph there is an agency with $\alpha =1$ if and only if there is a Hamiltonian cycle. 
\end{claim}

\noindent
{\bf Proof of Claim \ref{1}. }$\alpha =1$ implies $\alpha _2=1$, and thus $T=n$. The walk $a_1(0),a_1(0),$ $\cdots ,$ $a_1(T)$ is a Hamiltonian cycle. On the other hand, if there is a Hamiltonian cycle $v(i)$, then for the first agent we can set up a tour $a_1(t):=v(i)$, and all the other agents may follow this lead at a delay of $1,2, \cdots n-1$ time units. This makes a feasible agency. 

\medskip

\begin{claim}\label{7}
For any graph and any agency we have $\sqrt{T/k}\le \alpha \le T/k$. 
\end{claim}

\noindent
{\bf Proof of Claim \ref{7}. }
The claim follows from $\alpha _1 \, \alpha _2 = T/k$, and $1\le \alpha _1, \alpha _2\le \alpha$. 

\medskip

The main objective of this paper is to provide some nontrivial lower and upper bounds on the strength of a feasible agency. 

\section{Trees}

\begin{thm}\label{5}
If $G$ is a tree, then for any feasible agency we have $T/k \ge 4$. 
\end{thm}

\begin{claim} \label{c}
In any tree $G$ with $n\ge 3$ there is an edge $uv$ such that in $G-uv$ the component that contains $v$ is a star that is centered at $v$.
\end{claim} 

(A star is a $K_{1,m}$, where $m\ge 1$, and the node that corresponds to the color class of a single node is called its center. For ease of discussion, $K_2$ is called a star, and either of its nodes may be called its center.) 

\medskip

\noindent
{\bf Proof of Claim \ref{c}. }
To prove this claim, consider a longest path $P$ in the tree, and let $u,v,z$ be the last three nodes in this path. Edge $uv$ is a node as required: the component of $G-uv$ containing $v$ does have another node, namely $z$, connected to $v$. If some other node $z'$ in this component were to be not connected to $v$, then a path $P'$ ending in $z'$ would be longer than $P$, a contradiction. This proves the claim.

\medskip

\noindent
{\bf Proof of Theorem \ref{5}. }
We consider an edge $uv$ obtained from Claim \ref{c}, and let us denote the star by $v,z_1,z_2,\cdots ,z_m$, with $v$ its center. Let $U:=\{u,v,z_1, \cdots , z_m\}$. We distinguish three cases, depending on the value of $m$: $m=1$, $m=2$, and $m\ge 3$. 

\medskip

\begin{figure}
  \caption{Case $m\ge 3$.}
  \centering
  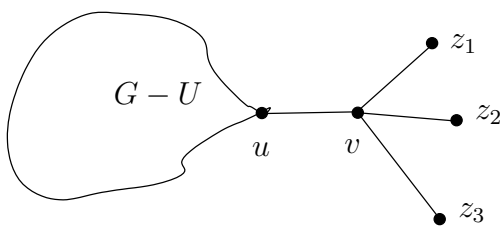
\end{figure}

\medskip

In case $m\ge 3$, the degree of node $deg_G(v)=m+1\ge 4$. Every agent needs to return at least $deg_G(v)$ number of times to $v$, and because they must not crash, $k \, deg_G(v)\ge T$. This implies $T/k\ge m+1\ge 4$. 

\medskip

\begin{figure}
  \caption{Case $m=1$.}
  \centering
  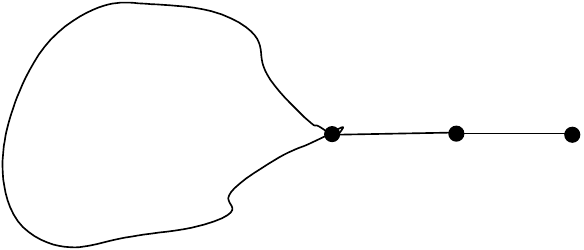
\end{figure}

\medskip

In case $m=1$, we are mainly concerned with node $z_1$. Every agent must visit $z_1$ at least once. Now two different agents visiting node $z_1$, say at times $t,t'$, respectively, that is, $a_i(t)=a_j(t')=z_1$. We claim that $|t-t'|\ge 5$. To prove this, it is quite easy to see that if $|t-t'|\le 4$, then agents $i,j$ would crash at a time between $t,t'$ in a node or edge in $G[\{u,v,z_1\}]$, a contradiction.
Thus $|t-t'|\ge 5$, and actually, this still is true if we were to repeat the walks of the agents after time $T$, modulo $T$. This implies that $T\ge 5k$.

\medskip

\begin{figure}
  \caption{Case $m=2$.}
  \centering
  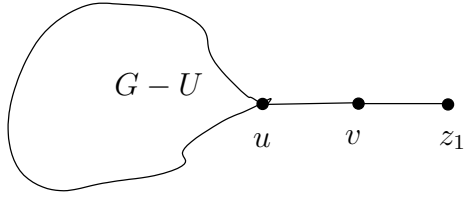
\end{figure}

\medskip

In case $m=2$ we need to be somewhat more careful to prove $T/k\ge 4$. Every agent must visit $z_1$ and $z_2$ at least once, and for each $i$ we consider the time interval(s) that agent $i$ spends in the nodes $U:=\{u,v,z_1,z_2\}$ while visiting at least one of the nodes $z_1$ or $z_2$. This means that for every agent $i$ we pick one or two intervals so that they include one visit to $v_1$ and one visit to $v_2$. For an agent $i$ there will be one or two corresponding intervals, say $I=[t^1, t^2]$. We consider time intervals modulo $T$, or "cyclically". Of course $I$ will start and end in a time unit when agent $i$ is occupying node $u$. Actually, we may assume that these time units satisfy the following: 

\begin{itemize}
\item
$a_i(t^1)=a_i(t^2)=u$, and 
\item
$a_i(s)\in \{z_1, z_2\}$ for some $s\in [t^1, t^2]$
\end{itemize}

By removing parking in $u$ from the beginning and end of the time interval $(t^1, t^2)$, we may also assume that 

\begin{itemize}
\item
$a_i(t^1+1)=a_i(t^1-1)=v$ 
\end{itemize}

\begin{defn}
An interval that satisfies these 3 properties for agent $i$ is called a $U$-interval for agent $i$. 
\end{defn}

Basically, a $U$-interval is the time interval that corresponds to a segment of that walk of agent $i$ that looks like $u,v,\cdots , z_1,\cdots ,v,u$ (or $u,v,\cdots , z_2,\cdots ,v,u$) such that this segment of the walk stays within node subset $U$. The definition requires that during this time interval agent $i$ starts with $u,v$, ends with $v,u$, and it touches $z_1$ or $z_2$ at least once, while staying within $U$. 
Of course there may be time intervals when an agent visits $u$, or even $v$, but without touching $z_1,z_2$ -- we just don!t call such intervals $U$-intervals. 

Every agent $i$ must visit $z_1$ and $z_2$ at least once, and thus we can determine 1 or 2 $U$-intervals for agent $i$ such that 
\begin{itemize}
\item
if there is one $U$-interval for agent $i$, than both $z_1$ and $z_2$ are visited by agent $i$ during this $U$-interval $I$, or 
\item
if there are two $U$-intervals for agent $i$, than $z_1$ is visited by agent $i$ during only one of these $U$-intervals, and $z_2$ is visited by agent $i$ during only the other $U$-interval. The two $U$-intervals for agent $i$ are disjoint. 
\end{itemize}
Let $\mathcal{I}_i$ denote the set of $U$-intervals for agent $i$, and let let $\mathcal{I}:=\bigcup _i\mathcal{I}_i$ denote the set of all $U$-intervals for all agents.  
The most important property of $U$-intervals is that no 3 of them overlap at the same time unit, or, in other words, $\mathcal{I}$ is a 2-packing. 

\begin{claim}\label{c1}
 For any $0\le t\le T$, there are at most two intervals of $\mathcal{I}$ that contain $t$, that is, $|\{I: I\in \mathcal{I}, t\in I\}|\le 2$. 
\end{claim}

To prove this, suppose there are three different agents for which the corresponding intervals contain $t$. This means that when the last of these agents enters $U:=\{u,v,z_1,z_2\}$, the two others are already in $U$. Say the last to enter $U$ is agent $i$, and the other two agents are $i', i''$. Say $t^1$ is the time when $i$ enters $U$. Agent $i$ at time $t^1+1$ moves to $v$, which implies that agent $i',i''$ occupy nodes $z_1,z_2$. By our assumption about these time intervals, at some time agent $i$ while staying inside $U$, will to move to $z_1$ or $z_2$ -- but the way is blocked by agents $i',i''$. A crash would be inevitable, and this contradiction proves the Claim. 

\medskip

The size of a time interval $I$ is measured by how many time units it has, that is, $|I\cap \mathbb{Z}|$. For short we
abbreviate this as $|I|:=|I\cap \mathbb{Z}|$. 
We claim that for any agent $i$, if there is one $U$-interval for agent $i$, then that time interval has at least 7 time units; and if there are two intervals for agent $i$, then both of these time intervals contain at least 5 time units. 

Formally, 
\begin{claim}\label{c2}
For any agent $i$, 
\begin{enumerate}
 \item 
if $\mathcal{I}_i=\{I\}$, then $|I|\ge 7$, and 
 \item 
  if $\mathcal{I}_i=\{I, J\}$, then $|I|, |I|\ge 5$, and 
\end{enumerate}
\end{claim}

To prove this, not that the shortest walks to visit $z_1$ and satisfy the properties required for a $U$-interval is $u,v,z_1,v,u$, which requires 5 time units. And the shortest walks to visit $z_1$ and $z_2$ and satisfy the properties required for a $U$-interval is $u,v,z_1,v,z_2,v,u$, which requires 7 time units. This proves Claim \ref{c2}. 

\medskip

By Claim \ref{c1} the set of intervals is a 2-packing, and this $2T\ge \sum _{I\in \mathcal{I}}|I|$. By Claim \ref{c2} we get that $\sum _{I\in \mathcal{I}}|I|\ge 7k$. This implies that $T/k\ge 3.5$. But our goal is to improve this bound to 4 from 3.5, so we need to look at these intervals more carefully. We have to look carefully especially at agents $i$ for which there is only one interval, and that interval contains 7 time units. 
  
The idea is to count tokens associated with time units. For every time unit $0\le t\le T-1$ we create 2 tokens, this is $2T$ tokens in total. 
We would 
\begin{itemize}
 \item 
award one of these tokens to an agent $i$ such that $t\in \bigcup \mathcal{I}_i$ 
\end{itemize}
Because of Claim \ref{c1}, we are not running out of tokens. By Claim \ref{c2}, each agent has received at least 7 tokens, proving that $2T\ge 7k$. This comes just short of proving $T/k\ge 4$, so we need to be a bit more careful with the assignment of tokens. To improve on this bound, we adjust the placement of tokens as follows: 
\begin{itemize}
 \item 
If only one token is taken from time unit $t$ by agent $i$, but the second token is up for grabs, then we award the second token to agent $i$, too. We do this for all time units $t$ like this. 
 \item
If $I\in \mathcal{I}_i, J\in \mathcal{I}_j$, such that $I\subseteq J$ and $|I|=7$, then agent $j$ gives one of its tokens to agent $i$. We do this between all pairs $I,J$ like this. 
\end{itemize}
\begin{claim}\label{c8}
 All agents have at least 8 tokens. 
\end{claim}

To prove this claim, we first show the following property of the intervals. Essentially, this claims that any $U$-interval $I$ of 7 time units either has a time unit which overlaps with no other $U$-interval of any other agent, or there a $U$-interval of another agent contains $I$ as a subset. 

\begin{claim}\label{c3}
If $\mathcal{I}_i=\{I\}$ with $|I|=7$, then exactly one of the following assertions must hold: 
\begin{enumerate}
 \item 
There is a time unit $t\in I - \bigcup _{J\in \mathcal{I}-I}J$. 
 \item
There is another interval $J\in \mathcal{I}-I$ such that $I\subseteq J$. 
\end{enumerate}
\end{claim}

In other words, Claim \ref{c3} states that for every agent with a single visit to $U$ that takes the minimum 7 time units, there is either a time unit when $i$ is the only agent on a visit to $U$, or there must be another agent that will on a visit in $U$ through these whole 7 time units. Actually, this means that this other agent needs an interval of length at least 9 to visit $U$: at least one time unit before and at least one after $i$'s visit. 

\medskip

\noindent
{\bf Proof of Claim \ref{c3}. }
The only way for this claim to be false would be if there were (at least) two other intervals, say $J_1$ and $J_2$, which overlap with $I$, and so that there is no gap between $J_1,J_2$. Say $J_1$ corresponds to agent 1, $J_2$ corresponds to agent 2. Let us denote these $U$-intervals by $I=[a,a+6]$, $J_1=[b,c]$ and $J_2=[c+1,d]$, and thus by our assumption $a\le c<c+1\le a+6$. This makes the choice of $c=a,a+1,a+2,a+3,a+4,a+5$ possible. Note that by the definition of a $U$-interval, agents 1 and 2 occupy nodes $v,u,u,v$ for time units $c-1,c,c+1,c+2$. Also by the definition of a $U$-interval, agent $i$ occupies nodes $u,v, \star , \star, \star, v,u$ at time units $a,a+1,a+2,a+3,a+4,a+5,a+6$. In all 6 possible cases of $c$, a crash happens in node $u$ or $v$, or edge $uv$. This contradiction proves Claim \ref{c3}.

\medskip

\noindent
{\bf Proof of Claim \ref{c8}. }
An agent $i$ with two $U$-intervals has at least 10 tokens, because in both of its $U$-intervals it received at least 5 tokens. Say one of its $U$-intervals is $J$. If the agent has had to give one of them for another agent by our rule above for some $I\subseteq J$, then it has kept the 6 other tokens from the time units in $I$. In this case agent $i$ has at least 6 tokens from $U$-interval $J$. Otherwise agent $i$ has at least 5 tokens from $U$-interval $J$. So this makes at least 10 tokens for this agent, considering both of its $U$-intervals. 

Consider an agent $i$ with one $U$-interval $J$ such that $|J|\ge 8$. If $i$ has not had to give away any of its tokens then we are done. So now assume that agent $i$ gave up one of its tokens because of another agent with a $U$-interval $I\subseteq J, |I|=7$. In this case $J$ must have at least one time unit before $I$, and at least one time unit after $I$, making it $|J|\ge 9$. So agent $i$ has received at least 9 tokens, and gave away only one of them. If $J$ overlaps with multiple  other $U$-intervals $I, |I|=7$ in this way, then it will keep at least 6 tokens from each of them, this way agent $i$ will have at least $6+6=12$ tokens in the end. Considering all cases, agent $i$ will have at least 8 tokens in the end. 

Finally, consider an agent $i$ with one $U$-interval $I$ such that $|I|=7$. This agent receives 7 tokens, and by Claim \ref{c3} it gets at least one more token for a time unit $t$ where agent $i$ is alone, or one token from another agent that is responsible for a $U$-interval $J$ such that $I\subseteq I$. In both cases, this agent will hold at least 7 tokens in the end. This proves Claim \ref{c8}. 

\medskip

Initially there are $2T$ tokens, and in the end we have accounted for at least $8k$ of them. This proves $T/k\ge 4$, and thus Theorem \ref{5}. 

\medskip

By Claim \ref{7} this also implies a lower bound on the strength of a feasible agency that we might get in a tree. 

\begin{cor}\label{c5}
If $G$ is a tree, then for any agency we have $\alpha \ge 2$. 
\end{cor}

The following example shows that the bound $T/k\ge 4$ given in Theorem \ref{5} is tight, for certain trees. This example, though, is not tight for the bound given in Corollary \ref{c5} -- we conjecture that that bound could be improved.

\medskip

\noindent
{\bf Example 1.}

We define a tree $G=(V,E)$ as follows. Consider a tree that has $3r+2$ (where $r$ is some positive integer) nodes such that $r$ of them have degree $4$ and $2r+2$ of them have degree 1. Let $F$ denote this set of nodes of degree 4, and let $L$ denote this set of $2r+2$ nodes of degree 1. We add another $4r+4$ nodes so that every node in $L$ gets 2 new neighbors: say if $v\in L$, then we add nodes $v^1$ and $v^2$ joined by an edge each to node $v$. Let us say $M$ denotes the set of new nodes. This defines a tree $G=(V,E)$ with $V=F\cup L\cup M$, having $7r+6$ nodes and $7r+5$ edges.

\begin{figure}
\centering
\begin{subfigure}{.7\textwidth}
  \centering
  \scalebox{1.0}{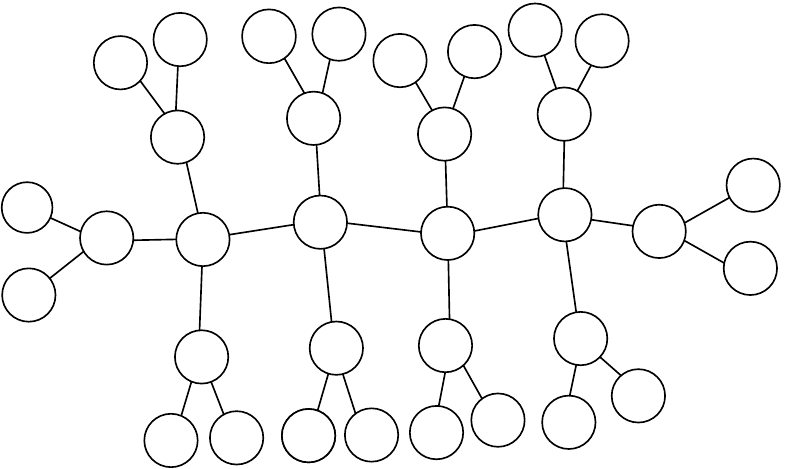}
  \caption{Time $t=t_0$.}
\end{subfigure}%

\bigskip

\begin{subfigure}{.7\textwidth}
  \centering
  \scalebox{1.0}{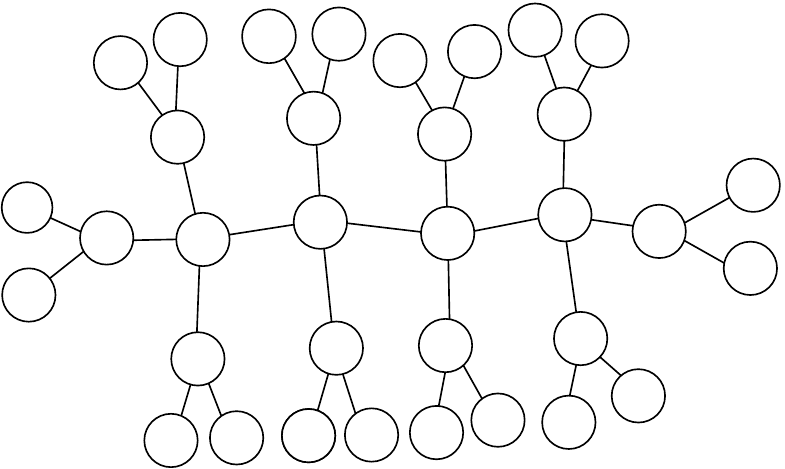}
  \caption{Time $t=t_0+1$.}
\end{subfigure}

\bigskip

\centering
\begin{subfigure}{.7\textwidth}
  \centering
  \scalebox{1.0}{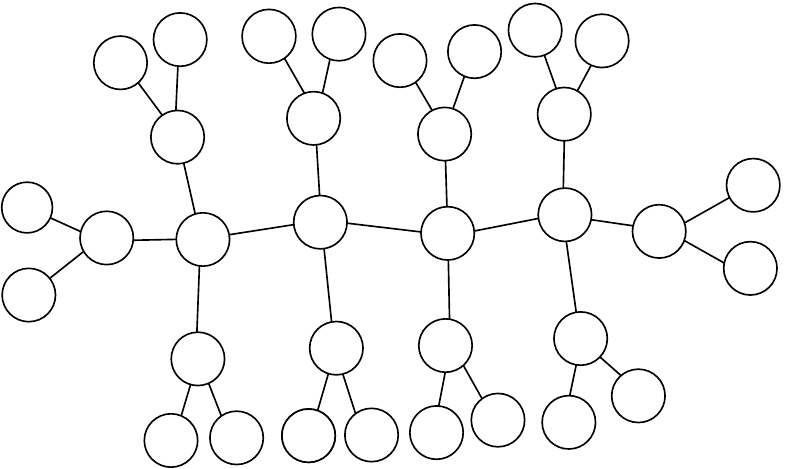}
  \caption{Time $t=t_0+2$.}
\end{subfigure}%

\bigskip

\begin{subfigure}{.7\textwidth}
  \centering
  \scalebox{1.0}{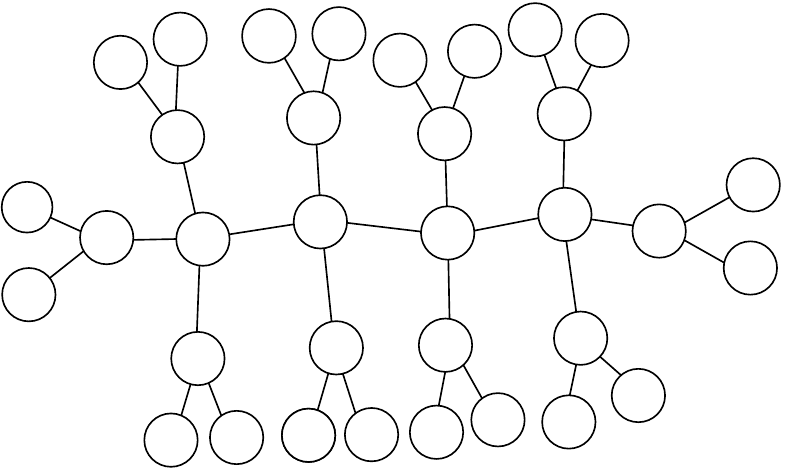}
  \caption{Time $t=t_0+3$.}
\end{subfigure}%

\caption{Example 1. Annotated by time units modulo 4.}
\label{fig:ex1}
\end{figure}

\bigskip

Consider an arbitrary closed walk of $G$ that visits every nodes at least once, traverses every edge exactly twice, and returns to its origin in $14r+10$ time units. We modify this walk so that for every node $v\in L$, we add parking in $v_1$ for one time unit -- a change that provides a walk that spends 2 time units in $v_1$. This change creates a walk of time horizon $T=(14r+10)+(2r+2)=16r+12$. Let us denote this walk by $v(t)$. For agent 1, we define $a_1(t):=w(t)$. For all other agetns, we define their walks $a_i$ by repeating $w$ at a delay of $4(i-1)$, repeated cyclically, that is $a_i(t):=w(t-4(i-1))$ where time is measured modulo $T$, i.e.\ cyclically. This defines a set of $T/4=4r+3$ agents. 

It is relatively straightforward to verify that this agency is feasible, thus proving the following claim. (For nodes $v\in L$ it can be verified manually that there is no crash in nodes or edges in $G[\{v, v^1,v^2\}]$. To see that there is no crash on any other node we need to observe that for any edge $e$ adjacent with at least one node in $F$, any agent spends 3 modulo 4 time units in either component of $G-e$ -- this implies that there is no crash in nodes of $F$ or edges adjacent with a node in $F$.) 

\begin{claim}
 There are trees of arbitrarily large size with a feasible agency such that $T/k=4$ (with $T=\frac{16}{7}n-\frac{12}{7}$ and $k=\frac{4}{7}n-\frac{3}{7}$).
\end{claim}

\section{Trees with $T=2n-2$ }

When graph $G$ is a tree, then any traveling salesman tour has time horizon at least $2m-2$, and thus of course, any agency has $T\ge 2n-2$. A tour with $T=2n-2$ is shortest possible, meaning that every edge of the tree is traversed only twice. In this section we restrict ourselves to shortest tours, that is $T=2n-2$, and our main result shows an upper bound on the number of agents in this case. 

\begin{thm}\label{4}
If $G$ is a tree, then for a feasible agency with $T=2n-2$ we have $T/k \ge 5$. 
\end{thm}

\medskip

\noindent
{\bf Proof of Theorem \ref{4}. }

Consider an edge $uv$ as guaranteed by Claim \ref{c}, and let us denote the star by $v,z_1,z_2,\cdots ,z_m$. We distinguish three cases, depending on the value of $m$, $m=1$, $m=2$, and $m\ge 3$. Let $U:=\{v,z_1,z_2,\cdots ,z_m\}$. (Let us remark that we consider the tours to be repeating cyclically modulo $T$, that is, we consider time units "modulo $T=2n-2$".) 
As in the proof of Theorem \ref{5}, every agent has a $U$-interval: for agent $i$ let the shortest time interval that contains the times when agent $i$ visits some $z_j$, and starts and ends with a time unit when agent $i$ is occupying node $u$ is called a $U$-interval. Because the tour of any agent $i$ is shortest, there is only one $U$-interval, and it has $2m+3$ time units. Say for agent $i$, the $U$-interval is equal to $[t(i), t(i)+2m+2]$. The walk of agent $i$ during its $U$-interval looks like $u,v,\cdots ,v,u$, and contains all leaves $z_j$ in $U$. 

\bigskip

Case $m=1$. In this case it takes $2m+3=5$ time units for any agent $i$ to complete its $U$-interval, which is $[t(i),t(i)+4]$. The segment of its tour during this $U$-interval must be $(u,v,z_1,v,u)$. It is plain and easy to see that the $U$-intervals of no two agents may overlap, to avoid a crash in $U$. Thus at least 5 time units need to be spent before another agent may start its own $U$-interval. 

\begin{claim}\label{c9}
If $i\ne j$ then $[t(i), t(i)+4]\cap[t(j), t(j)+4]=\emptyset$. 
\end{claim}

Now, for all time units between 1 and $T$ we create one token. An agent takes all tokens corresponding to its $U$-interval. Because of Claim \ref{c9} $5k$ tokens are distributed among the agents. This proves the desired bound $T\ge 5k$ in this case. 

\bigskip

Case $m=3$. In this case it takes $2m+3=9$ time units for agent $i$ tocomplete its $U$-interval, which is $[t(i),t(i)+8]$. The segment of its tour during this $U$-interval must be $(u,v,z_a,v,z_b,v,z_c,v,u)$, where $a,b,c$ is a permutation of $1,2,3$. To avoid a crash in node $v$, if two agents $i,j$ have overlapping $U$-intervals, then the intersection of their $U$-interval must be even. 

\begin{claim}\label{c17}
If $i\ne j$, then $[t(i), t(i)+4]\cap[t(j), t(j)+4]$ has even cardinality. 
\end{claim} 

This claim implies that any given time unit $t$ may be part of at most two $U$-intervals corresponding to any agents, or we may say the $U$-intervals are a 2-packing. (Considering just this, we obtain a bound of $9k\le 2T$.) Actually, this also implies the following: 

\begin{claim}\label{c18}
For any agent $i$ there is a time unit $t$ in its $U$-interval (i.e.\ $t\in [t(i),t(i)+8]$) such that $t$ is not part of any other agent's $U$-interval. 
\end{claim}

Otherwise there would have to be two other agents, $i'$ and $i''$ such that $[t(i), t(i)+8]\subseteq [t(i'), t(i')+8]\cup [t(i''), t(i'')+8]$. Because a $U$-interval has an odd number of time units, $t(i')$ and $t(i'')$ need to have different parity. This contradicts Claim \ref{c17}, and thus proves Claim \ref{c18}. 

We now use Claims \ref{c17} and \ref{c18} to finish the proof in this case. We create 2 tokens for every time unit $1\le t\le T$. The token corresponding to time unit $t$ is awarded to agent $i$ if $t\in [t(i),t(i)+8]$. Because the $U$-intervals are a 2-packing, the tokens created are enough for this step. Now if for $t$ only one token is taken, say by agent $i$, then the other token of $t$ is also awarded to agent $i$. Because a $U$-interval has 9 time units,  every agent will have 9 tokens are taken initially. Because of Claim \ref{c18}, every agent will have at least one extra token, thus every agent will receive at least 10 tokens in total. This implies $2T\ge 10k$, proving the bound in this case. 

Case $m>3$ is solved the same way. 

\bigskip

Case $m=2$.In this case it takes $2m+3=7$ time units for agent $i$ to complete its $U$-interval, which is $[t(i),t(i)+6]$. The segment of its tour during this $U$-interval must be $(u,v,z_a,v,z_b,v,u)$, where $a,b$ is a permutation of $1,2$. To avoid a crash in node $v$, if two agents $i,j$ have overlapping $U$-intervals, then the intersection of their $U$-interval must be even. 

\begin{claim}\label{c21}
If $i\ne j$, then $[t(i), t(i)+6]\cap[t(j), t(j)+6]$ has even cardinality. 
\end{claim}

Moreover, to avoid a crash on edge $uv$, the overlap $[t(i), t(i)+6]\cap[t(j), t(j)+6]$ may not have cardinality 2. To avoid a crash in edges $vz_1, vz_2$, the overlap  $[t(i), t(i)+6]\cap[t(j), t(j)+6]$ may not have cardinality 6. This implies that when there is an overlap, it must have cardinality exactly 4. Because the $U$-intervals are a 2-packing, this implies the following:

\begin{claim}\label{c22}
For any agent $i$ there is are 3 different time units $t$ in its $U$-interval (i.e.\ $t\in [t(i),t(i)+8]$) such that $t$ is not part of any other agent's $U$-interval. 
\end{claim}

We now use Claims \ref{c21} and \ref{c22} to finish the proof in this case. We create 2 tokens for every time unit $1\le t\le T$. The token corresponding to time unit $t$ is awarded to agent $i$ if $t\in [t(i),t(i)+8]$. Because the $U$-intervals are a 2-packing, the tokens created are enough for this step. Now if for $t$ only one token is taken, say by agent $i$, then the other token of $t$ is also awarded to agent $i$. Because a $U$-interval has 7 time units, every agent will have 7 tokens are taken initially. Because of Claim \ref{c18}, every agent will have at least three extra token, thus every agent will receive at least 10 tokens in total. This implies $2T\ge 10k$, proving the bound in this case.

This completes the proof of Theorem \ref{4}.

\medskip

\begin{cor}\label{c4}
If $G$ is a tree, then for any agency with $T=2n-2$ we have $\alpha \ge \sqrt{5}$. 
\end{cor}

\noindent 
{\bf Example 2.} 
The following example shows that the bound $T/k\ge 5$ given in Theorem \ref{4} is tight, for certain trees. This example, though, is not tight for the bound given in Corollary \ref{c4} -- we conjecture that that bound could be improved. We construct a graph as shown in the figure by using a path of even length $2q$, and adding a 1-path to its $2i+1$'th node, and adding a 2-path to its $2i+2$'th node. We add another 2-path to its first node, and a 3-star at its last node. This creates a graph of $n=5q+6$ nodes, and $5q+5$ edges. We create a walk of length $10q+10$ so that it first moves from left to right along the path, and also enters and leaves the 1-paths, then the walk enters and leaves the 3-star and all of its leaves, and then on the way back from right to left, the walk enters and leaves the 2-paths. This walk is indicated with the time units modulo 5 in the figure. Agent 1 would follow this tour, and agent $i$ would follow this tour at a delay of $5(i-1)$, cyclically. So that in time unit 1, the nodes indicated with "1" are occupied, and then after each step, the nodes indicated with "$j$" are occupied at time units congruent with $j$ modulo 5. It is quite easy to verify that this creates a feasible agency without parking with time horizon $T=10q+10$ and number of agents $k=2q+2=T/5$.

\begin{figure}
\centering
\begin{subfigure}{.7\textwidth}
  \centering
  \scalebox{.9}{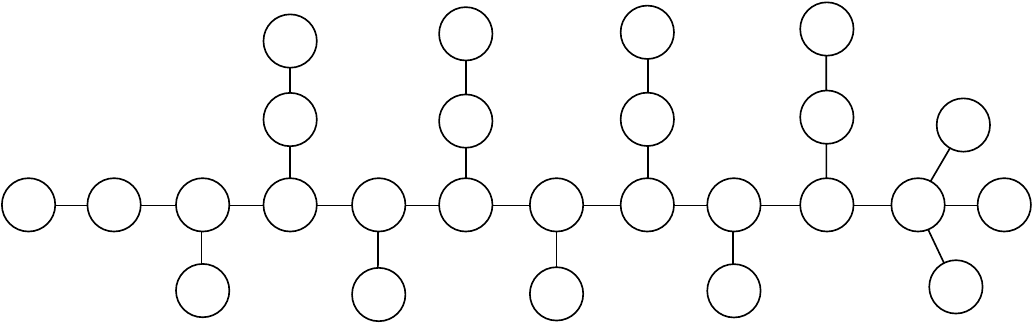}
  \caption{Time $t=t_0$.}
\end{subfigure}%

\bigskip

\begin{subfigure}{.7\textwidth}
  \centering
  \scalebox{.9}{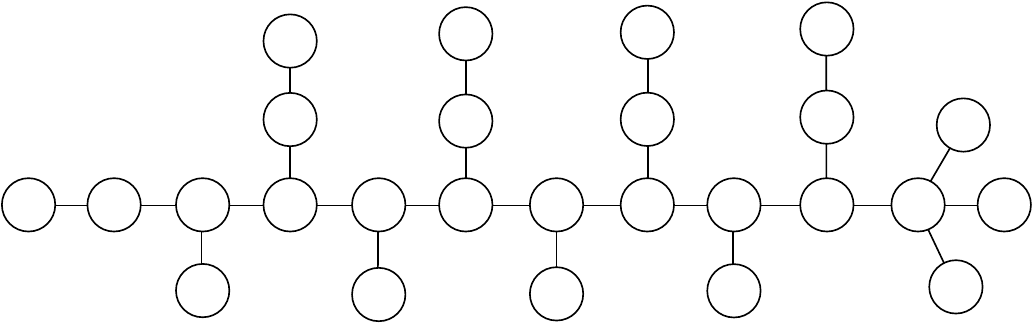}
  \caption{Time $t=t_0+1$.}
\end{subfigure}

\bigskip

\centering
\begin{subfigure}{.7\textwidth}
  \centering
  \scalebox{.9}{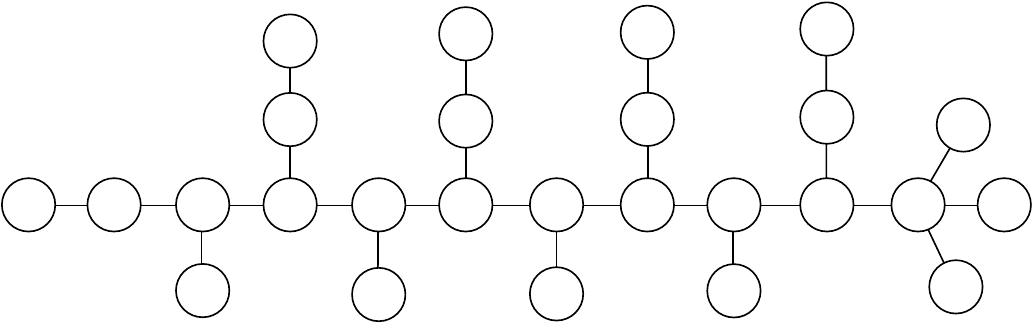}
  \caption{Time $t=t_0+2$.}
\end{subfigure}%

\bigskip

\begin{subfigure}{.7\textwidth}
  \centering
  \scalebox{.9}{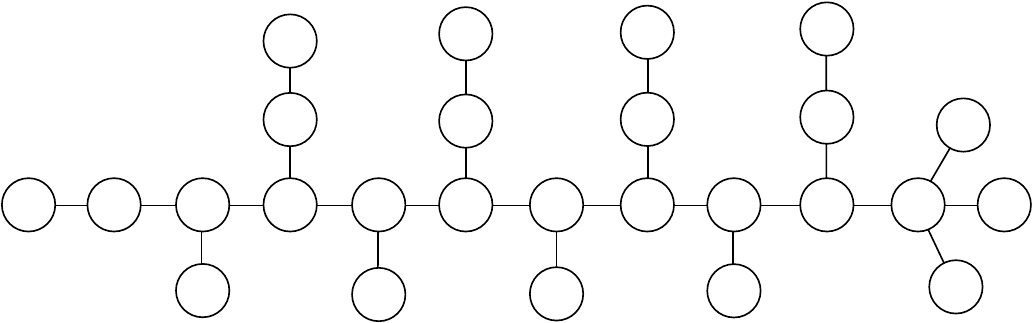}
  \caption{Time $t=t_0+3$.}
\end{subfigure}%

\bigskip

\centering
\begin{subfigure}{.7\textwidth}
  \centering
  \scalebox{.9}{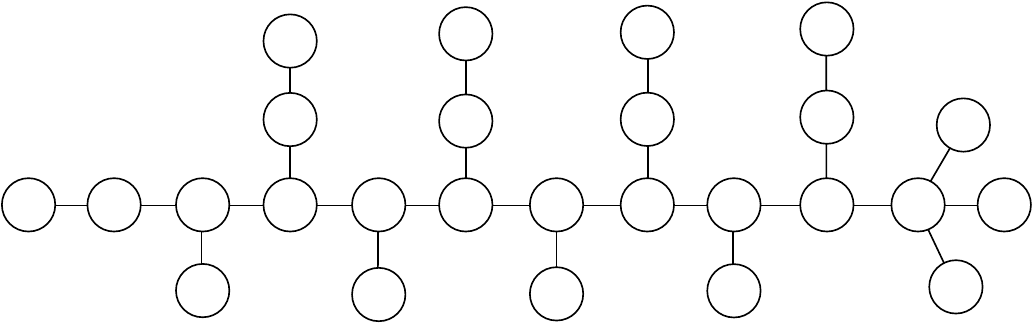}
  \caption{Time $t=t_0+2$.}
\end{subfigure}%

\caption{Example 2. Annotated by time units modulo 5.}
\label{fig:ex2}
\end{figure}

\bigskip

\begin{thm}\label{2}
 There are arbitrarily large trees for which there is a feasible agency without parking such that $T=2n-2$ and $k=\frac{1}{5}T$. 
\end{thm}

\section{3-edge-connected 3-regular graphs}

\begin{thm}\label{3}
 If $G=(V,E)$ is a 3-regular 3-edge-connected graph, then there is an agency with $T/k=4$, $T=2n, k=n/2$ and $\alpha =\alpha_1=\alpha _2=2$. 
\end{thm}
 
By Petersen's theorem we know that graph $G$ contains a perfect matching, or equivalently, by taking the complement of a perfect matching, we also know that $G$ contains a 2-factor (a 2-regular subgraph spanning all the nodes). There are ways of formulating a result that is stronger than Petersen's Theorem, and one of them is a recent result of Boyd, Iwata and Takazawa \cite{bit}, claiming the existence of a 2-factor of a special property.

\begin{thm}[Boyd, Iwata, Takazawa\cite{bit}]\label{bit}
 If $G=(V,E)$ is a 3-regular bridgeless graph, then there is a 2-factor covering every 3-cut and every 4-cut. 
\end{thm}

\medskip

\noindent
{\bf Proof of Theorem \ref{3}. }
Because of Theorem \ref{bit}, there is a perfect matching $M$ in $G$ such that $G-M=(V,E-M)$ consists of cycles of length at least 5. Let $\mathcal{C}=\{C_1,C_2,\cdots ,C_s\}$ denote the set of these cycles. Let $N\subseteq M$ be a spanning tree in $G/\mathcal{C}$. Thus $F:=N\cup \bigcup \mathcal{C}$ is connected, and the only cycles in $F$ are those of $\mathcal{C}$. A tour $a(t)$ of subgraph $(V,F)$ is obtained so that edges in $N$ are traversed twice, and edges in $\bigcup \mathcal{C}$ are traversed once (an arbitrary orientation of the cycles is chosen). The time horizon for this tour is $T=|V-V(N)|+4|N|$, because it uses both nodes of an edge in $N$ twice.  

We modify this tour $a(t)$ just by adding parking times to nodes in $V-V(N)$. We will come up with a "parking vector" $p:V-V(N)\rightarrow \{1,2,3,4\}$, and we interpret this with the intention of making an agent stay in a node $v\in V-V(N)$ for $p(v)$ time units. This means the agent will park for $p(v)-1$ time units. This defines a walk $a^p(t)$, which is obtained from $a(t)$ by repeating $v$ by $p(v)$ copies. The time horizon of this walk will be 
$$T=4|N|+\sum _{v\in V-V(N)}p(v).$$
Though this defines the tour of an agent for an arbitrary parking vector $p$, we need $p$ to satisfy a certain property in order to define a feasible agency. 
\begin{defn}
We say that the parking vector $p$ is a {\bf 4-cyclic parking vector} if for every $C\in \mathcal{C}$, the sum 
$$\sum _{v\in V(C)-V(N)}p(v)$$
is a multiple of 4. 
\end{defn}
The point of this definition is that a 4-cyclic parking vector may be used to define an agency. Let $k:=T/4$, and for $i=1,2,\cdots ,k$, let $a^p_i$ be the tour obtained from $a^p$ at a delay of $4(i-1)$. Thus all agents will have a traveling salesman tour, and it is also quite straightforward that this defines a feasible agency. 
\begin{claim}\label{c15}
For any 4-cyclic parking vector $p$, the set of tours $a^p_i$ defined above is a feasible agency. 
\end{claim}
To prove this claim, consider any edge $uv\in N$, which of course is a cut edge in $(V,F)$, say it separates the node set into parts $u\in V_u$ and $v\in V_v$. Also let us suppose that $t$ is the time unit when our agent comes to $u$ to move to $v$ in the following time unit, that is, $a^p(t)=u$ and $a^p(t+1)=v$. Because the number of time units our agent spends in $V_v-v$ is a multiple of 4, the time $t'$ when it returns to $v$ (i.e.\, $t'>t, a^p(t')=v$) is congruent with $t+2$ modulo 4, that is, $t'\equiv t+2 \, (mod\,  4)$. This implies that there is no crash in nodes $u$ or $v$ or on edge $uv$. Thus there is no crash on nodes or edges of matching $N$, but this also implies that there is no crash on nodes $V-V(N)$, because $p(v)\le 4$, and thus there is no time for agent $a^p_{i+1}$ to catch up with agent $a^p_i$. This proves Claim \ref{c15}

Of course, $\overline{p}(v):=4$ for all $v$ is a 4-cyclic parking vector, and thus it defines a feasible agency. Because of \ref{bit}, $|\mathcal{C}|\le \left\lfloor  n/5 \right\rfloor$, and thus $|N|\le \left\lfloor  n/5 \right\rfloor -1$. 
Then for $\overline{P}$ we get a feasible agency with 
$$T=4|N|+4|V-V(N)|=2|V|+2|V-V(N)|\ge 2n + 2(n-2(\left\lfloor  n/5 \right\rfloor -1))\ge \frac{16}{5}n.$$

Next we construct a different agency $a^p$ from a different 4-cyclic parking vector $p$, trying to make $p$ as small as possible. For a cycle $C_i\in \mathcal{C}$, the minimum number of time units spent in the nodes of $V(C)-V(N)$ just depends on $|V(C)-V(N)|$: we would want to have as many ones as possible, and then with the last remaining element, we fix the remainder modulo 4. Here we need the function $r(s):=4\left\lceil \frac{1}{4}s\right\rceil$, thus the minimum number number of time units spent in the nodes of $V(C)-V(N)$ is equal to $r(|V(C)-V(N)|)$. We get the following claim. 
\begin{claim}
There is 4-cyclic parking vector $p$ such that for agency $a^p$ the time horizon is equal to $T=4|N|+\sum _{C\in \mathcal{C}}r(|V(C)-V(N)|)$. 
\end{claim}
We will be able to prove a bound on $T$ by partitioning $\mathcal{C}=\mathcal{C}_0\cup \mathcal{C}_1\cup \mathcal{C}_2\cup \mathcal{C}_3$ such that 
\begin{itemize}
 \item 
$|V(C)-V(N)|=0$ for cycles $C\in \mathcal{C}_0$, 
 \item 
$|V(C)-V(N)|=1$ for cycles $C\in \mathcal{C}_1$, 
 \item 
$|V(C)-V(N)|=2$ for cycles $C\in \mathcal{C}_2$, 
 \item 
$|V(C)-V(N)|\ge 3$ for cycles $C\in \mathcal{C}_3$. 
\end{itemize}
By the definition of function $r(\cdot )$ we get that 
\begin{itemize}
 \item 
$r(|V(C)-V(N)|)=0=2|V(C)-V(N)|$ for $C\in \mathcal{C}_0$,
 \item 
$r(|V(C)-V(N)|)=4\le 2|V(C)-V(N)|+2$ for $C\in \mathcal{C}_1$,
 \item 
$r(|V(C)-V(N)|)=4=2|V(C)-V(N)|$ for $C\in \mathcal{C}_2$,
 \item 
$r(|V(C)-V(N)|)\le 2|V(C)-V(N)|-2$ for $C\in \mathcal{C}_3$.
\end{itemize}
Consider the tree $F:=(V,N)/\mathcal{C}$ that we get from edges in $N$ by shrinking the cycles of $\mathcal{C}$. All cycles in $\mathcal{C}_1$ correspond to a node of $F$ of degree at least 3. All leaves of $F$ correspond to a cycle in $\mathcal{C}_3$. 
For any tree, the number of leaves is larger than the number of nodes of degree at least 3, which for tree $F$ implies that $|\mathcal{C}_3|\ge |\mathcal{C}_1|$. We put all this together in the following calculation: 

{\small 

$$ 
T=4|N|+\sum _{C\in \mathcal{C}}r(|V(C)-V(N)|)=
$$

\begin{multline*}
 =2|V(N)|+\sum _{C\in \mathcal{C}_0}r(|V(C)-V(N)|)+\sum _{C\in \mathcal{C}_1}r(|V(C)-V(N)|)+
 \\
 +\sum _{C\in \mathcal{C}_2}r(|V(C)-V(N)|)+\sum _{C\in \mathcal{C}_3}r(|V(C)-V(N)|)\le 
\end{multline*}

\begin{multline*}
\le 2|V(N)|+\sum _{C\in \mathcal{C}_0}2|V(C)-V(N)|+\sum _{C\in \mathcal{C}_1}(2|V(C)-V(N)|+2)+
\\
+\sum _{C\in \mathcal{C}_2}2|V(C)-V(N)|+\sum _{C\in \mathcal{C}_3}(2|V(C)-V(N)|-2)=
\end{multline*}

$$
=2|V(N)|+\sum _{C\in \mathcal{C}}2|V(C)-V(N)|+ 2|\mathcal{C}_1| -2|\mathcal{C}_3| \le 2|V(N)|+\sum _{C\in \mathcal{C}}2|V(C)-V(N)|= 2|V|=2n.
$$

Thus we have constructed two different 4-cyclic parking vectors: $\overline{p}\equiv 4$ with $T\ge \frac{16}{5}n$, and this latter 4-cyclic parking vector $p$ with $T\le 2n$. We can add any multiple of 4 time units to $p$ while maintaining its 4-cyclic property, to obtain 4-cyclic parking vectors $p'$ with $p\le p'\le \overline{p}$, and thus we can find one that provides a feasible agency with $T=2n$. This proves Theorem \ref{3}.

}

\section{Optimization problems}
Optimization problems with regard to $k, T, \alpha , \alpha _1, \alpha _2$ seem to be rather hard to solve exactly -- recall that by Theorem \ref{1}, minimizing $\alpha$ is NP-hard. There are a few problems, though, that may be solved exactly, namely the following: 

\begin{problem}\label{p7}
For a given graph $G=(V,E)$, find a feasible agency (with parking) of $k=n=|V|$ agents and any time horizon $T$, or determine that there is no such agency. 
\end{problem}

\begin{problem}\label{p8}
For a given graph $G=(V,E)$, find a feasible agency without parking of $k=n=|V|$ agents and any time horizon $T$, or determine that there is no such agency. 
\end{problem}

We are going to prove that both of these problems can be solved in polynomial time. 

\begin{thm}\label{t7}
There is a polynomial time algorithm to solve Problem \ref{p7}. 
\end{thm}

To prove this, note that all nodes are occupied by an agent at any time. Thus the step between time unit $t$ and $t+1$ corresponds with a set of node-disjoint cycles in $G$. Agents in any of these cycles will move to their respective neighbors, cyclically, while agents in a node not covered by any of these cycles are just parking there during this time unit. This suggests that the following condition may describe the existence of a a feasible agency $k=n$ as required in Problem \ref{p7}: 

\begin{claim}\label{cl7}
There is a feasible agency with parking of $k=n$ agents if and only if $G$ is 2-edge-connected. 
\end{claim}

To show this claim, first note that if there is a cut edge $e$, then no agent may traverse $e$ because $e$ is not part of any cycle. Thus agent to one side of $G-e$ will never be able to get to the other side. To prove the other way around, suppose $G$ is 2-edge-connected. It is well-known that in this case, every edge is contained in a cycle. Now consider a spanning tree of $G$, and agent 1: we make agent 1 move along edges of the spanning tree, while completing each step of agent 1 using a cycle to define steps for all other agents. We repeat this for edges of the spanning tree until agent 1 has visited all the nodes. Then we repeat this for all other agents, one by one. In the end we need to repeat all these steps backwards to get all the agents back to their original position. The time horizon for this agency will be $T\le 2n(2n-3)$. This proves Theorem \ref{t7}.

\begin{conj} \label{conj8}
 There is a polynomial time algorithm to solve Problem \ref{p8}. 
\end{conj} 

To prove this, note that all nodes are occupied by an agent at any time. Thus the step between time unit $t$ and $t+1$ corresponds with a 2-factor in $G$. Recall that a 2-factor is a spanning subgraph such that all nodes have degree exactly 2. A 2-factor consists of cycles, and all the nodes are covered by exactly one cycle. This suggests that the following condition may describe the existence of a a feasible agency $k=n$ as required in Problem \ref{p8}: 

\begin{claim}\label{cl8}
There is a feasible agency without parking of $k=n$ agents if and only if the edges of $G$ which are contained in a 2-factor form a connected spanning subgraph. 
\end{claim}

To show this claim, first let us assume that there is a partition $V=V_1\cup V_2$ into two disjoint nonempty subsets such that edges between $V_1,V_2$ are contained in no 2-factor. In this case, an agent that starts out in $V_1$ will never leave $V_1$, which implies that there is no agency with the required properties. To prove the other way around, suppose that there is a spanning tree $F\subseteq E$ such that all edges in $F$ are contained in a 2-factor. Then it is straightforward to construct an agency: we make agent 1 move along edges of $F$, while completing each step of agent 1 using a 2-factor to define steps for all other agents. We repeat this for edges of the spanning tree until agent 1 has visited all the nodes. Then we repeat this for all other agents, one by one. After $n(2n-3)$ time units every agent has finished with visiting all the nodes. To obtain a feasible agency, we need to make sure all the agents will get back to their original position, all at the same time (this is required by the definition of a feasible agency). Let us say it took $T_1\le n(2n-3)$ time units to finish the moves up to this point, and note that the current positions of the agents define a permutation of $V$, where node $v\in V$ is mapped to node $\phi (v)\in V$ iff the agent starting in $v$ has arrived at $\phi (v)$ after $T_1$ time units. We just have to repeat $\phi$ a few times until all agents get back to their original positions, that is, we are looking for a positive integer $q$ such that $\phi ^q = \text{id}$. Here the least common multiple of cycle length's of cycles in $\phi$ will do -- the trivial upper bound on this is $q\le n!$. This proves Claim \ref{cl8} with an exponential bound on the time horizon, $T=qT_1\le n(2n-3)n!$. Unfortunately this is not polynomial, and for this reason we fall short of proving Conjecture \ref{conj8} -- though we are able to determine a yes/no answer to the existence of such an agency. At this point it is unclear if the existence of an agency with very long, maybe exponential time horizon would actually also imply the existence of an agency with polynomial time horizon, thus leaving the proof of Conjecture \ref{conj8} open. 

\bigskip

We may consider the problem of maximizing the number of agents in a particular graph, without a limit on the time horizon. The problem becomes similar to a mechanical disentanglement puzzle. The following partial solution to this was found in informal discussions with L\'aszl\'o V\'egh, Amitabh Basu, and Daniel Dadush at an Oberwolfach workshop. 

\begin{problem}\label{p9}
Suppose that $G=(V,E)$ is a tree. Find a feasible agency (with parking allowed) of a maximum number of agents $k$. (Note that there is no constraint on time horizon $T$.) \end{problem}

\begin{thm}\label{cl9}
If $G=(V,E)$ is a tree with no nodes of degree 2, then the following cases determine the optimum for Problem \ref{p9}:
\begin{itemize}
 \item[a)] If $n=1$ or $n=2$, then the optimum is $k=1$. 
 \item[b)] If $n\ge 3$ and $G$ is a star, then the optimum is $k=n-2$. 
 \item[c)] Otherwise the optimum is $k=n-3$. 
\end{itemize}
\end{thm}
(A tree with no node of degree 2 may not have 3 nodes.)

\medskip

\noindent
{\bf Proof. }
Proof of part a) and b) is rather simple:

If $n=2$ then the tree is just a single edge. We can have a single agent moving between the two endpoints of this edge. There is no agency with 2 agents, because they would bump into each other along the edge or one of its endpoints. Thus the optimum is $k=1$. 

If $G$ is a star with a center, and $n-1$ leaves, then we can define an agency with $n-2$ agents as follows. We initiate with the $n-2$ agents in some of the leaves. Either one of the agents may then go to the vacant leaf, by passing through the center. By repeating this $2n-2$ times, every agent will have the chance to visit each of the leaves. This is an agency with $n-2$ agents. There is no agency with $n-1$ agents, because in with $n-1$ agents, each agents would be restricted to just one of the leaves and the center - all other leaves are blocked by the other agents. Thus the optimum is $k=n-2$. 

In case c), graph $G$ is a tree with no degree-2 nodes, such that there is an edge $uv\in E$ such that both components of $G-e$ have more than one node. 

For contradiction, assume that there is an agency with $k=n-2$ agents. 
We partition $V$ into the following parts: $V=V_1\cup V_2\cup V_3\cup V_4$ such that $V_2=\{u\}$, $V_3=\{v\}$, and $V_1\cup V_2$ is one of the components of $G-uv$, and $V_3\cup V_4$ is the other component of $G-uv$. All parts $V_i$ are nonempty. 

We may assume that the agency is initiated so that $u$ and $v$ are the two empty nodes, and all other nodes are occupied by agents. (We may assume this, because any initial setup could be reached from this state, or actually, any other state.) Let us "color" the agent so that agents that are in $V_1$ in the initial setting are colored red, and agents that are in $V_1$ in the initial setting are colored blue. We claim that red and blue agents will not mix, and actually, at all times the agent distribution will always be like one of the following cases: (1) red agents occupy a subset of $V_1\cup V_2\cup V_3$ and blue agents occupy all of $V_4$, or (2) red agents occupy a subset of $V_1\cup V_2$ and blue agents occupy a subset of $V_3\cup V_4$, or (3) red agents occupy all of $V_1$ and blue agents occupy a subset of $V_2\cup V_3\cup V_4$. This is easily seen because (a) $V_2$ and $V_3$ are sets of just a single node, and thus red and blue cannot pass by each other there, and furthermore, (b) when there is a blue agent in $V_2$ then all of $V_1$ is occupied, so this blue agent cannot enter $V_1$ to mix with the red agents there. Anyway, we conclude that red agents will not be able to visit nodes in the nonempty subset $V_4$, and blue agents will not be able to enter nodes in the nonempty $V_1$. Thus there is no agency of this size $k=n-2$. 

In case c), the tree must be of diameter at least 3, because it is not a star, and it also has no nodes of degree 2. To prove that there is a feasible agency with $k=n-3$, it suffices to prove the following claim: 

\begin{claim}\label{cl11}
Suppose there are $n-3$ agents, agent $i$ is currently in node $u\in V$, and there is an edge $uv\in E$. Then there is a sequence of agency moves such that agent $i$ finds itself in node $v$ after that sequence of moves. 
\end{claim}

\noindent
{\bf Proof. }
Let us denote by $V_u$ and $V_v$ the nodeset of the two components of $G-uv$, so that $V=V_u\cup V_v$, $u\in V_v$, $v\in V_v$. If there is an unoccupied node in $V_v$, then a sequence of moves along a path will result in $v$ being unoccupied, and then agent $i$ may move into node $v$ as required. Thus we may assume that all nodes in $V_v$ are occupied. This also implies that all 3 unoccupied nodes must be in $V_u$. 

Because $u$ is occupied (by agent $i$), the 3 unoccupied nodes are in $V_u-u$. $G[V_u]-u$ partitions into a family of components, say denoted by $H_1,H_2,\cdots , H_p$. Now we distinguish two cases here: either A) there are at least two distinct components say $H_1,H_2$ that contain at least one unoccupied node, or B) all 3 unoccupied nodes are in the same component, say $H_1$. 

\begin{figure}
\centering
\begin{subfigure}{.33\textwidth}
  \centering
  \scalebox{.7}{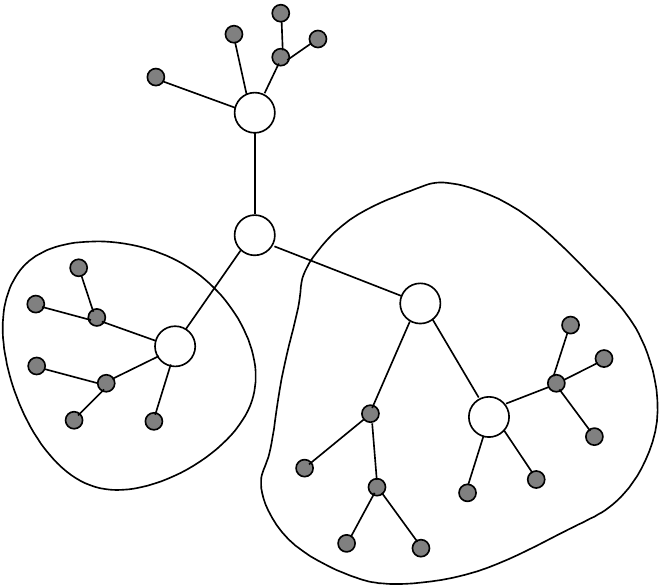}
  \caption{Time $t=t_0$.}
\end{subfigure}%
\begin{subfigure}{.33\textwidth}
  \centering
  \scalebox{.7}{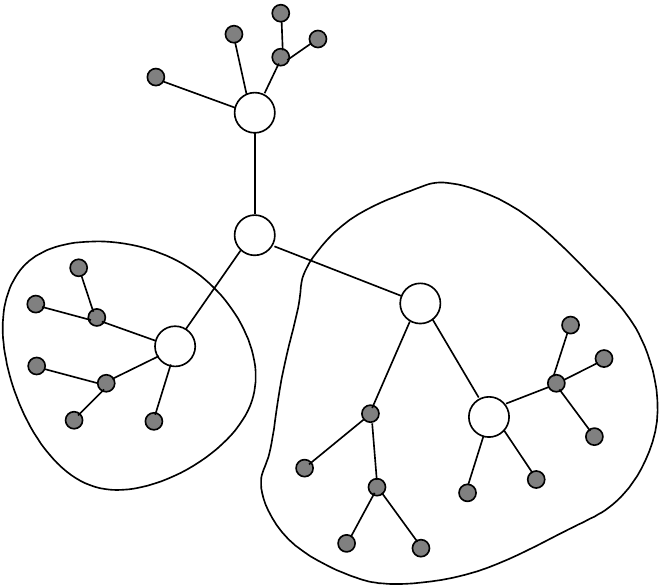}
  \caption{Time $t=t_0+1$.}
\end{subfigure}%
\begin{subfigure}{.33\textwidth}
  \centering
  \scalebox{.7}{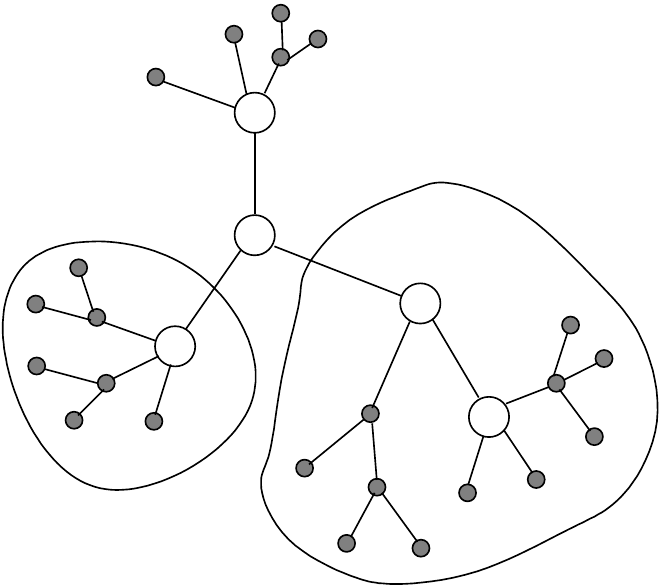}
  \caption{Time $t=t_0+2$.}
\end{subfigure}

\bigskip

\begin{subfigure}{.33\textwidth}
  \centering
  \scalebox{.7}{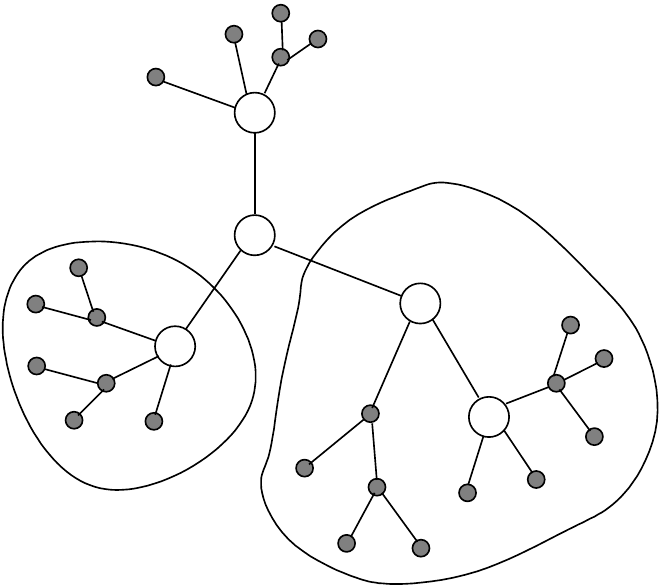}
  \caption{Time $t=t_0+3$.}
\end{subfigure}%
\begin{subfigure}{.33\textwidth}
  \centering
  \scalebox{.7}{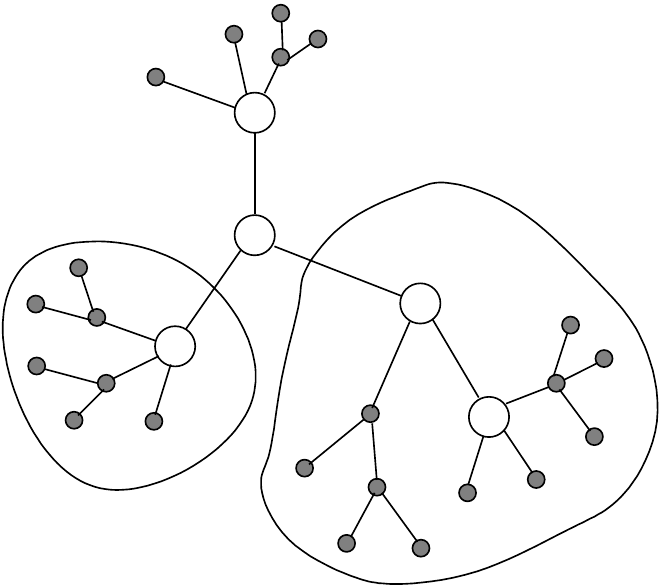}
  \caption{Time $t=t_0+4$.}
\end{subfigure}%
\begin{subfigure}{.33\textwidth}
  \centering
  \scalebox{.7}{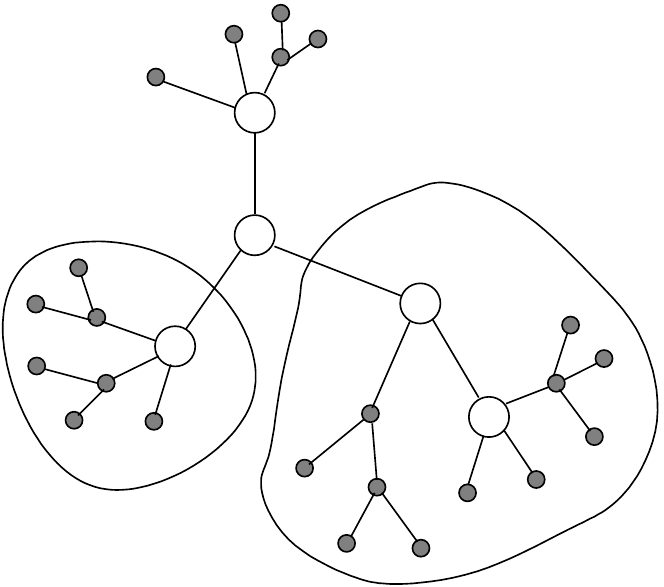}
  \caption{Time $t=t_0+5$.}
\end{subfigure}

\bigskip

\begin{subfigure}{.33\textwidth}
  \centering
  \scalebox{.7}{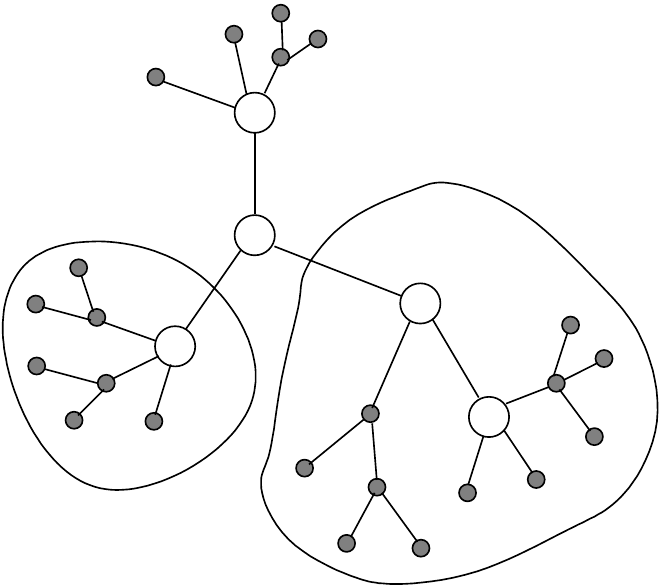}
  \caption{Time $t=t_0+6$.}
\end{subfigure}%

\caption{Case A in the proof of Claim \ref{cl11}.}
\label{CaseA}
\end{figure}

In case A), a simple sequence of moves will allow agent $i$ to be pushed into component $H_1$, because initially $H_1$ has contained an unoccupied node. Then $u$ becomes unoccupied, and thus a simple sequence of moves will allow agent $j$ to be pushed into $H_2$, because initially $H_1$ has contained an unoccupied node. After this, both $u$ and $v$ become unoccupied, and we may allow both agents $i$ and $j$ to return to nodes $u$ and $v$. We may, however, choose to first allow $i$ to reach node $v$, and then allow $j$ to return to node $u$. Anyway, the main observation is that after this sequence of moves, agent $i$ gets to visit node $v$ as we wanted to prove. An example of this action is show in in Figure \ref{CaseA}. 

\begin{figure}
\centering
\begin{subfigure}{.33\textwidth}
  \centering
  \scalebox{.7}{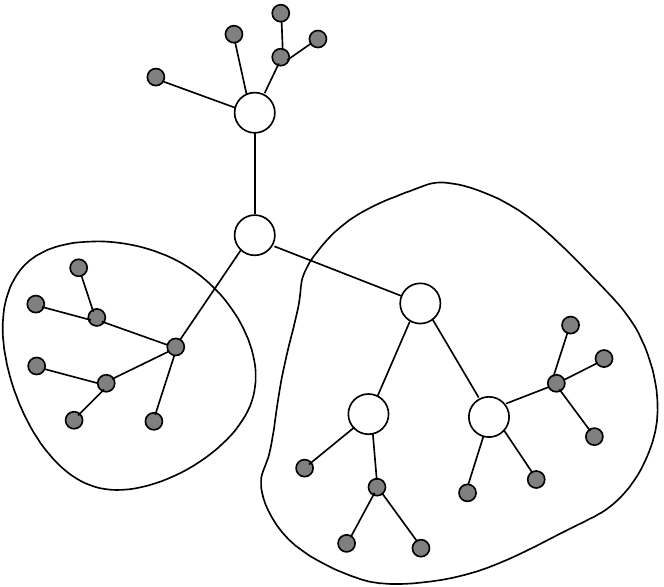}
  \caption{Time $t=t_0$.}
\end{subfigure}%
\begin{subfigure}{.33\textwidth}
  \centering
  \scalebox{.7}{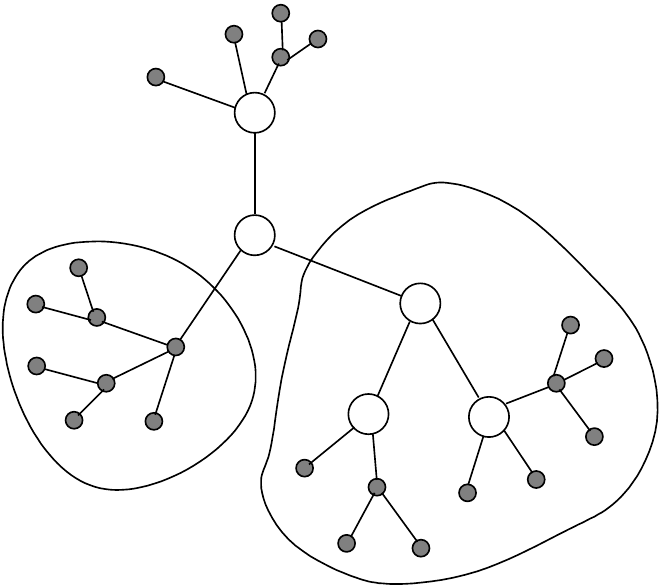}
  \caption{Time $t=t_0+1$.}
\end{subfigure}%
\begin{subfigure}{.33\textwidth}
  \centering
  \scalebox{.7}{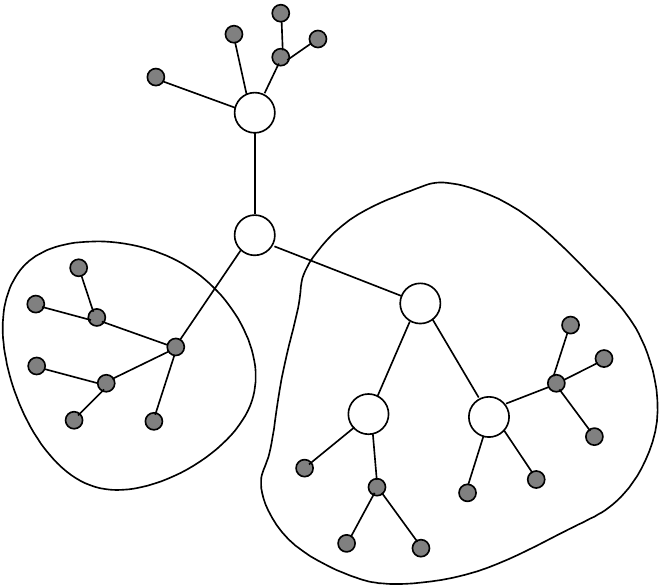}
  \caption{Time $t=t_0+2$.}
\end{subfigure}

\centering
\begin{subfigure}{.33\textwidth}
  \centering
  \scalebox{.7}{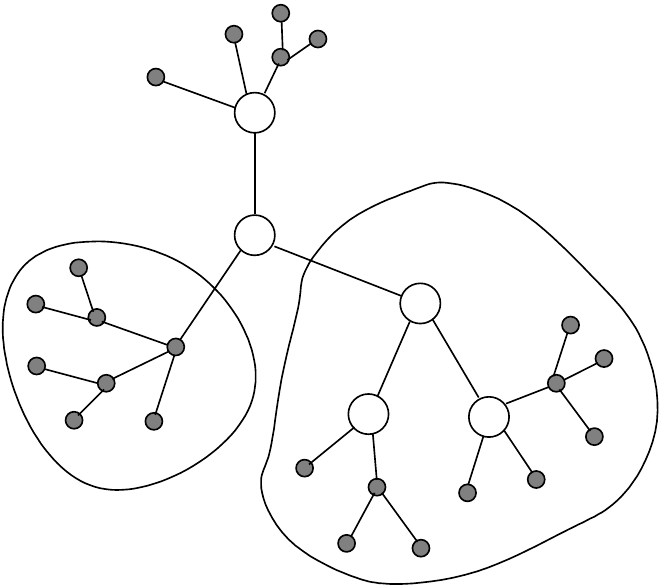}
  \caption{Time $t=t_0+3$.}
\end{subfigure}%
\begin{subfigure}{.33\textwidth}
  \centering
  \scalebox{.7}{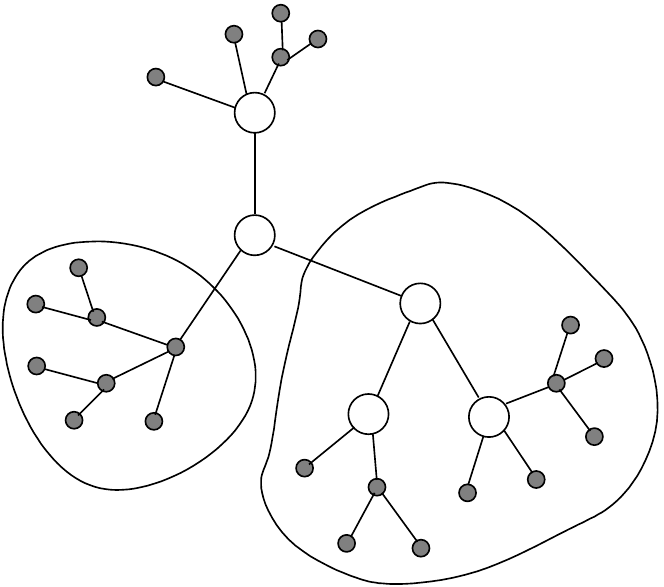}
  \caption{Time $t=t_0+4$.}
\end{subfigure}%
\begin{subfigure}{.33\textwidth}
  \centering
  \scalebox{.7}{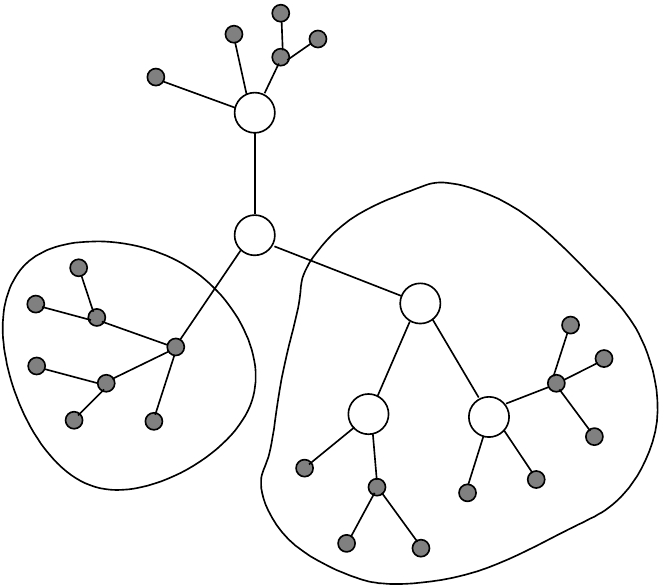}
  \caption{Time $t=t_0+5$.}
\end{subfigure}

\centering
\begin{subfigure}{.33\textwidth}
  \centering
  \scalebox{.7}{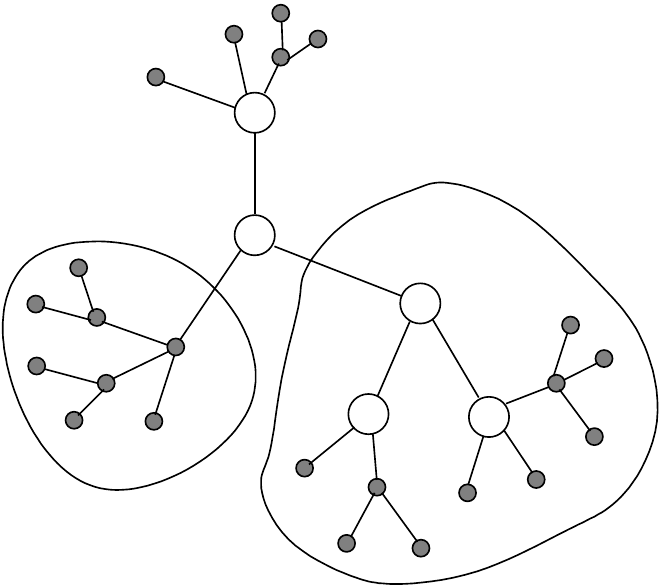}
  \caption{Time $t=t_0+6$.}
  \label{fig:sub1}
\end{subfigure}%
\begin{subfigure}{.33\textwidth}
  \centering
  \scalebox{.7}{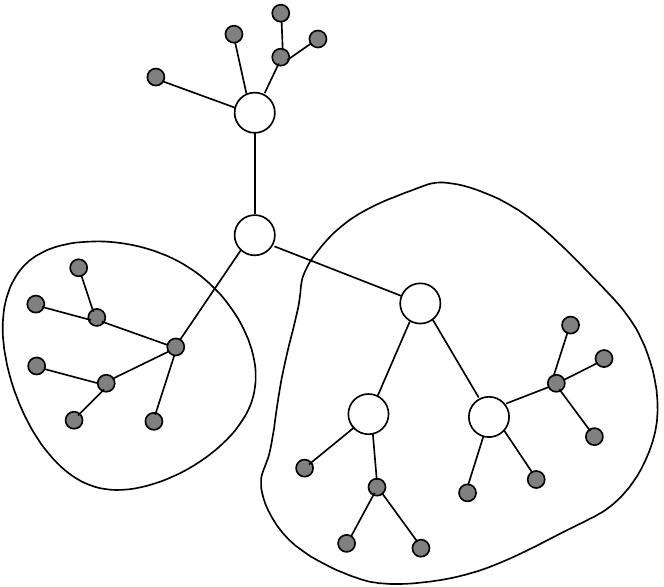}
  \caption{Time $t=t_0+7$.}
\end{subfigure}%
\begin{subfigure}{.33\textwidth}
  \centering
  \scalebox{.7}{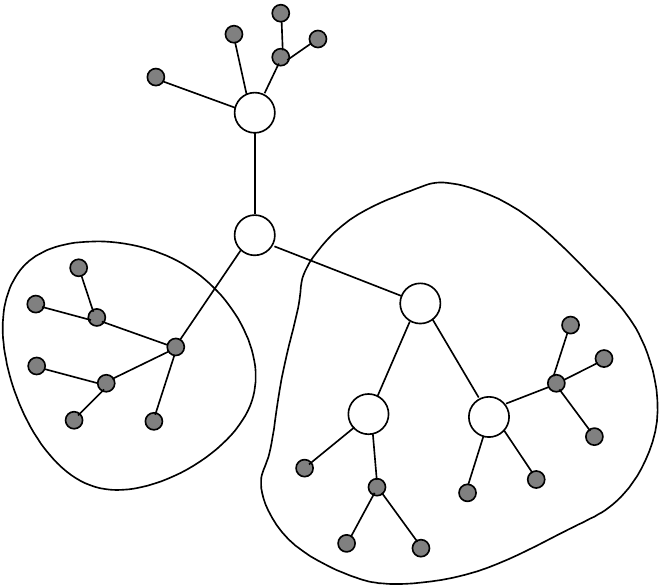}
  \caption{Time $t=t_0+8$.}
\end{subfigure}

\centering
\begin{subfigure}{.33\textwidth}
  \centering
  \scalebox{.7}{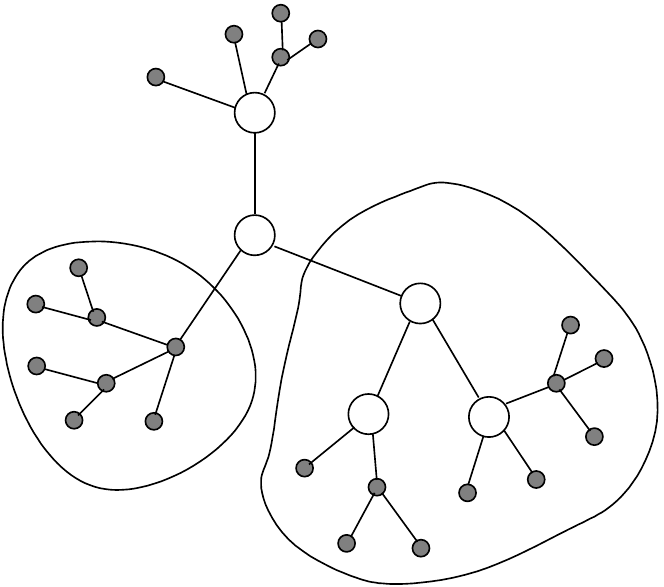}
  \caption{Time $t=t_0+9$.}
\end{subfigure}%
\begin{subfigure}{.33\textwidth}
  \centering
  \scalebox{.7}{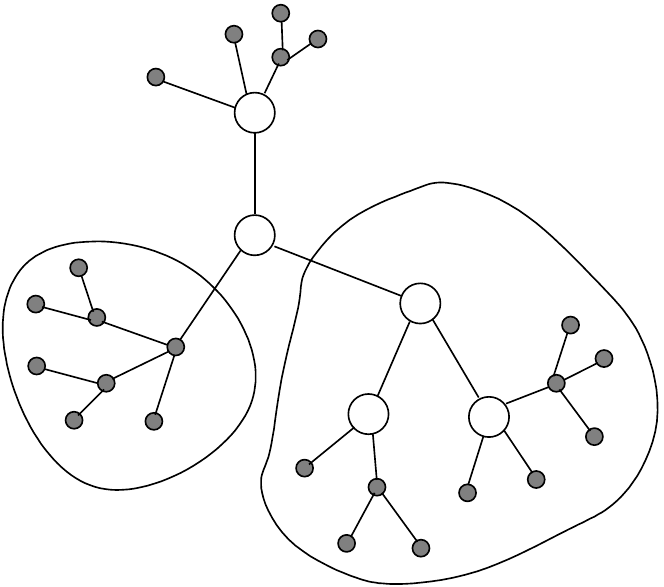}
  \caption{Time $t=t_0+10$.}
\end{subfigure}%

\caption{Case B in the proof of Claim \ref{cl11}.}
\label{CaseB}
\end{figure}

In case B), we have the situation shown in Figure \ref{CaseB}, where in component $H_1$ there is a node $u_1$, a neighbor of $u$, and $u_1$ has two more neighbors, $u_{11}$ and $u_{12}$. Agent $i$ may go to $u_{12}$ in two steps, and then agent $j$ can go to $u_{11}$ in two steps. After that, agent $i$ may go to $v$ in three steps, and agent $j$ may go to $u$ in two steps. After all these steps, agent $i$ has succeeded in going to $v$, as claimed.

\medskip

\bigskip

\begin{figure}
\centering
\begin{subfigure}{.33\textwidth}
  \centering
  \scalebox{.6}{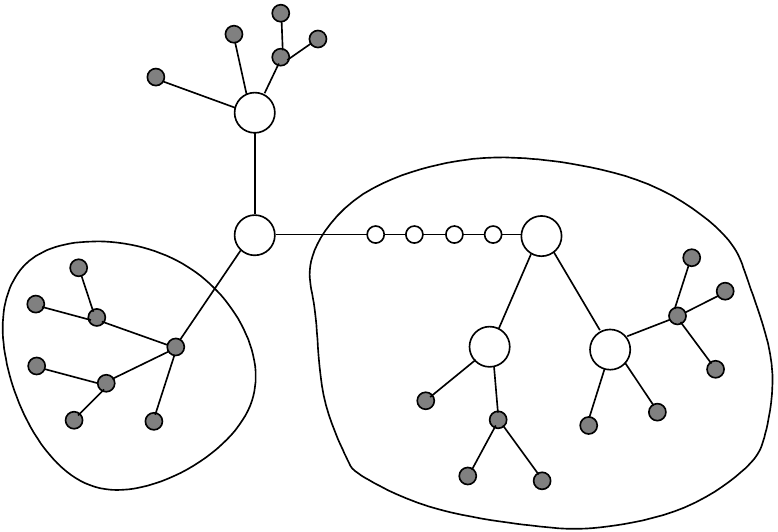}
  \caption{Time $t=t_0$.}
\end{subfigure}%
\begin{subfigure}{.33\textwidth}
  \centering
  \scalebox{.6}{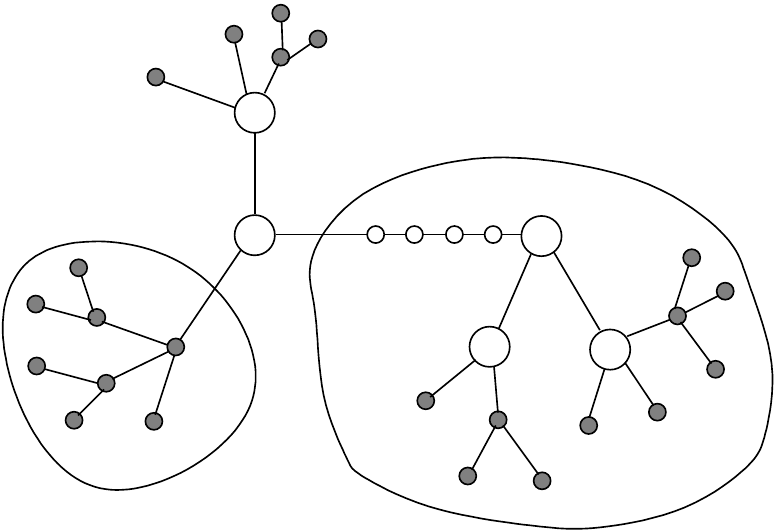}
  \caption{Time $t=t_0+\sigma $.}
  \label{fig:sub2}
\end{subfigure}%
\begin{subfigure}{.33\textwidth}
  \centering
  \scalebox{.6}{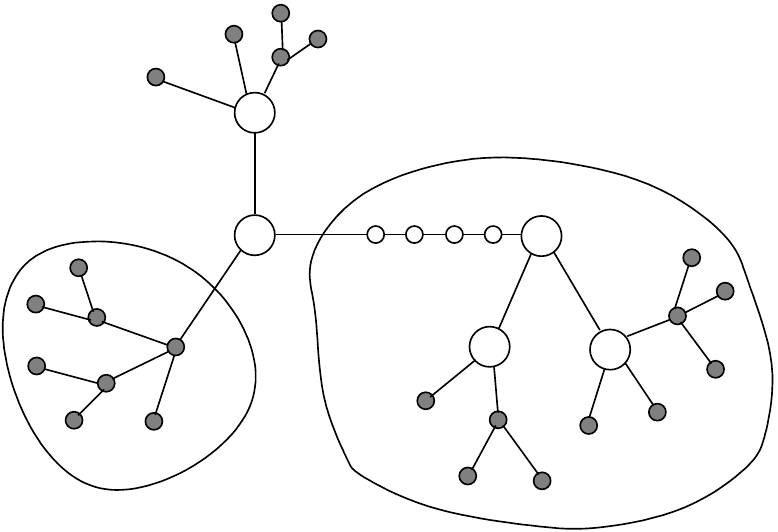}
  \caption{Time $t=t_0+\sigma +1$.}
\end{subfigure}

\bigskip

\centering
\begin{subfigure}{.33\textwidth}
  \centering
  \scalebox{.6}{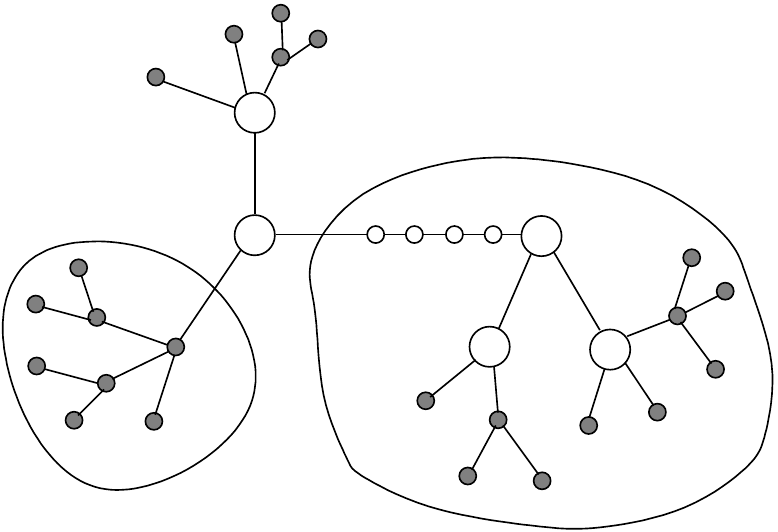}
  \caption{Time $t=t_0+\sigma +2$.}
\end{subfigure}%
\begin{subfigure}{.33\textwidth}
  \centering
  \scalebox{.6}{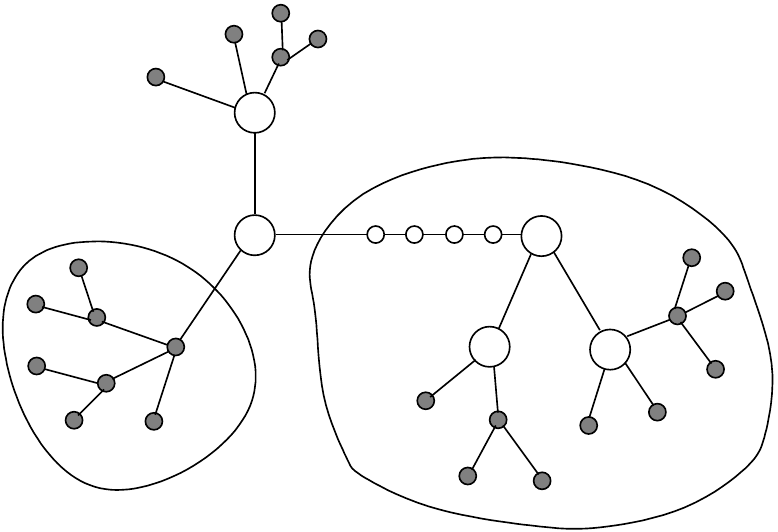}
  \caption{Time $t=t_0+2\sigma +2$.}
\end{subfigure}%
\begin{subfigure}{.33\textwidth}
  \centering
  \scalebox{.6}{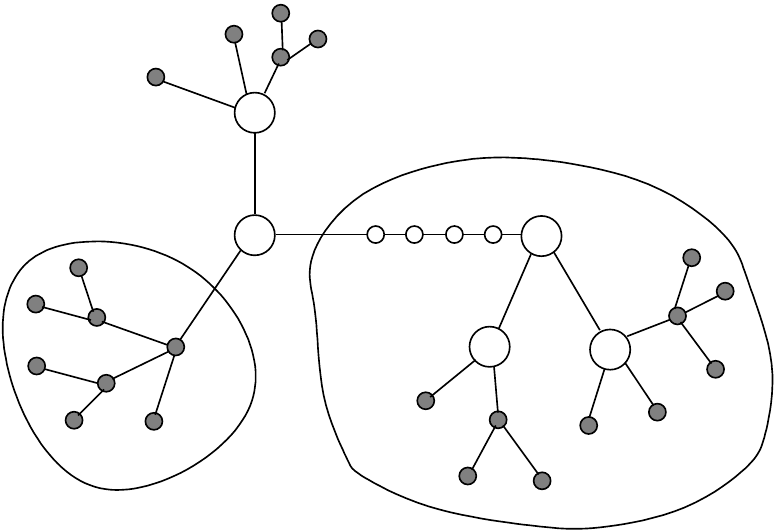}
  \caption{Time $t=t_0+2\sigma +3$.}
\end{subfigure}

\bigskip

\centering
\begin{subfigure}{.33\textwidth}
  \centering
  \scalebox{.6}{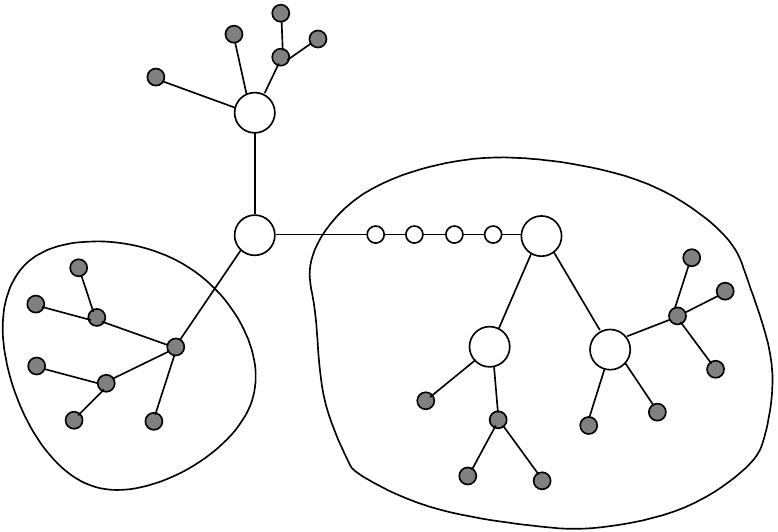}
  \caption{Time $t=t_0+2\sigma +4$.}
\end{subfigure}%
\begin{subfigure}{.33\textwidth}
  \centering
  \scalebox{.6}{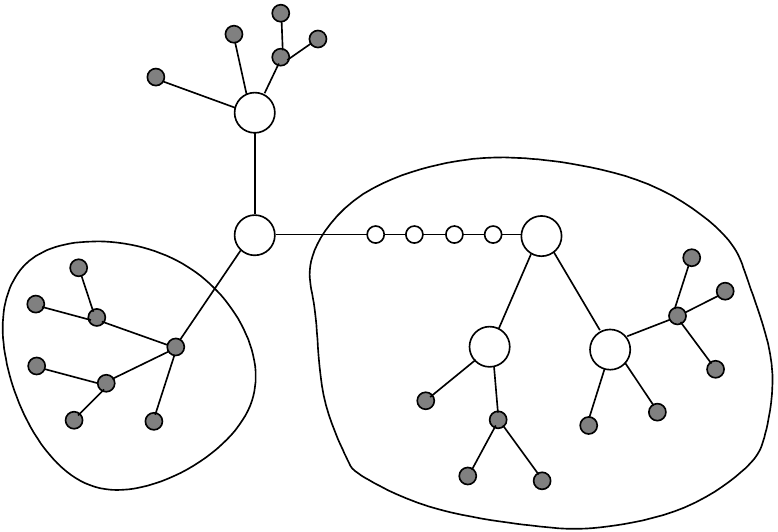}
  \caption{Time $t=t_0+3\sigma +4$.}
\end{subfigure}%
\begin{subfigure}{.33\textwidth}
  \centering
  \scalebox{.6}{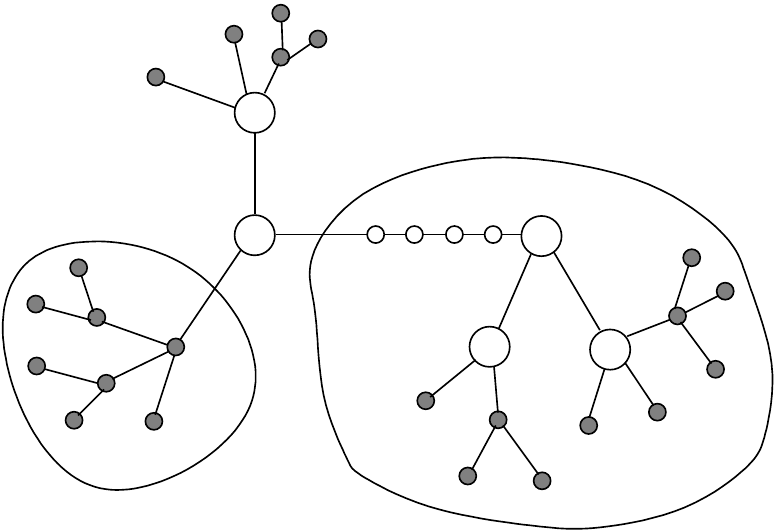}
  \caption{Time $t=t_0+3\sigma +5$.}
\end{subfigure}

\bigskip

\centering
\begin{subfigure}{.33\textwidth}
  \centering
  \scalebox{.6}{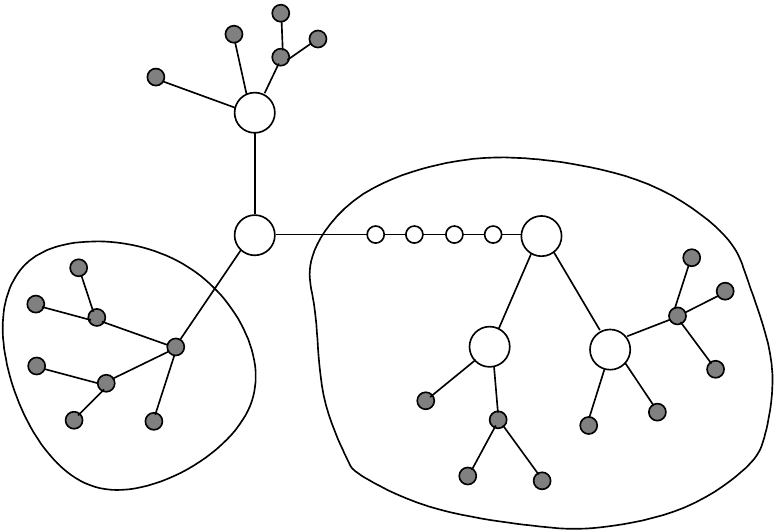}
  \caption{Time $t=t_0+3\sigma +6$.}
\end{subfigure}%
\begin{subfigure}{.33\textwidth}
  \centering
  \scalebox{.6}{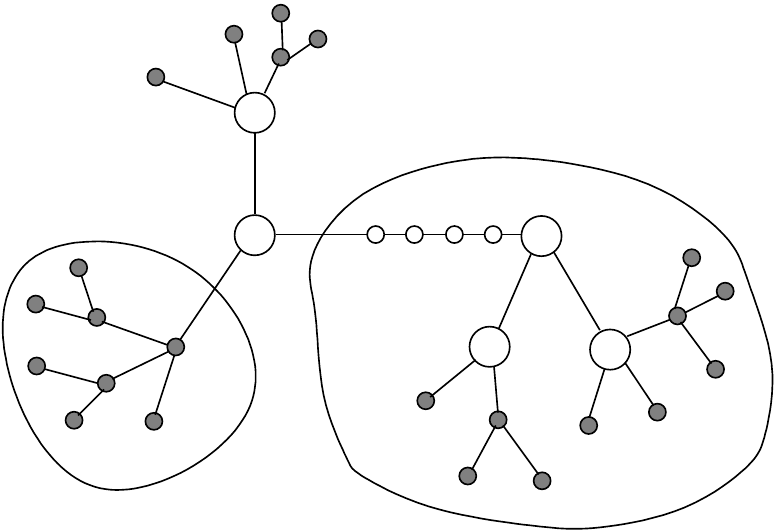}
  \caption{Time $t=t_0+4\sigma +6$.}
\end{subfigure}%

\caption{Case C in the proof of Claim \ref{cl15}. We denote $\sigma := dist_T(u,u_1)$}. 
\label{CaseC}
\end{figure}

\bigskip

In a tree $G=(V,E)$, let $L\subseteq V$ denote the set of leaves (nodes of degree 1), and let $Y\subseteq V$ denote the set of "joints", i.e.\ nodes with degree at least 3. An $L$--$Y$-path is a path in $G$ that connects a leaf with a joint, and apart from its endpoints, all points of the path have degree 2. An $Y$--$Y$-path is a path in $G$ that connects a joint with a joint, and apart from its endpoints, all points of the path have degree 2. For a tree that has at least one joint (i.e.\ is not a path), we define its \emph{stretch} as follows: 

$$stretch_{YY}(G):=
\left\{ 
\begin{aligned}
&\max \{|V(P)|: P \text{ is a $Y$--$Y$-path}\} &&\text{ if there is a $Y$--$Y$-path}
\\
&-\infty &&\text{ otherwise} 
\end{aligned}
\right.
$$

$$stretch_{LY}(G):=
\left\{ 
\begin{aligned}
&\max \{|V(P)|: P \text{ is an $L$--$Y$-path}\} &&\text{ if there is an $L$--$Y$-path}
\\
&-\infty &&\text{ otherwise} 
\end{aligned}
\right.
$$

$$stretch_{LL}(G):=\left\{ 
\begin{aligned}
&|V(P)| &&\text{ if $G=P$ is a path}
\\
&-\infty &&\text{ otherwise} 
\end{aligned}
\right.$$

\begin{thm}\label{cl15}
If $G=(V,E)$ is a tree with no nodes of degree 2, then the optimum for Problem \ref{p9} is equal to: 
$$k=\min \{n-stretch_{YY}(G)-1, n-stretch_{LY}(G), n-stretch_{LL}(G)+1\}.$$ 
\end{thm}

This theorem makes sense in the following special cases: When $G$ is a star, then from $stretch_{LY}=1$ we get that $k=n-2$, which is straightforward. When $G$ is a path, then from $stretch_{LL}=n$ we get that $k=1$, which is actually quite obvious for a path. When $G$ is a tree that has no nodes of degree 2, and more than 2 nodes, then from $stretch_{YY}=2$ we get that $k=n-3$, just like we expected from Theorem \ref{cl9}. 

\medskip

\noindent
{\bf Proof of Theorem \ref{cl15}. }
We first prove that $k$ is at most the value on the right hand side. cf

To see that $k\le n-stretch_{LL}(G)+1$ is rather trivial: If $G$ is not a path, then this value is infinity, and we are done. Otherwise, if $G$ is a path, then the value become equal to 1, which is an obvious upper bound on the number of agents in a path. 

To see that $k\le n-stretch_{LY}(G)$, assume that $G$ is not a path (because in case of a path, this value would be infinity, and we are done.) Consider the path $P$ where the maximum attains, let us say it is a path $P$ that connects leaf $l\in V$ with a node $v\in V$. Thus $dist_G(u,v)=stretch_{YY}(G)-1$.  All interior nodes of $P$ have degree 2 in $G$. At a point in time, the leaf $l$ will be occupied by an agent $i$, say. If for contradiction, $k\ge n-stretch_{LY}(G)+1$, then it would be impossible for agent $i$ to leave the path $P$, because as long as it stays in $P$, the part of $P$ between $l$ and the agent will be empty, and thus, when the agent gets to node $v$, all nodes of $G$ outside of $P$ will be occupied by the other agents. Thus agent $i$ would not be able to visit nodes outside of $P$, and because $P$ is not a path, this contradicts the definition of an agency. 

The proof of $k\le n-stretch_{YY}(G)-1$ will be similar to the proof of the upper bound in case c) of Theorem \ref{cl9}. Consider the $Y$--$Y$ path $P$ that attains maximum length, connecting nodes $u$ and $v$ of degree at least 3. Let $V_v$ denote the set of node in $V-v$ that are in the same component of $G-u$ as $v$. Similarly, let $V_u$ denote the set of node in $V-u$ that are in the same component of $G-v$ as $u$. Note that $dist_G(u,v)=stretch_{YY}(G)-1$. All interior nodes of $P$ have degree 2 in $G$. Now assume for contradiction, that $k=n-stretch_{YY}(G)$. This means that in the graph we have $|V(P)|$ number of unoccupied nodes. Note that for the purpose of this proof, we may assume that at $t=0$ the unoccupied nodes are exactly those of $V(P)$ (because from that state we may reach any other initial state that was supposed to happen). So there are two groups of agents, those agents near to $u$ occupying the nodes of $V_u$ (called $u$-agents), and those agents near $v$ occupying nodes of $V_v$ (called $v$-agents). Then it is quite easy to see that $u$-agents and $v$-agents "do not mix", that is, whatever moves the agents are doing, one of the edges incident with a node in $V(P)$ cuts away all $u$-agents from all $v$-agents. Note that this property is preserved by induction, because a $u$-agent can only make moves in $V-V_v$, and a $v$-agent can only make moves in $V-V_u$. Because of this, when for example a $u$-agent gets as far as node $v$, then all nodes of $V_v$ will be occupied by $v$-agents, and thus this agent will have to stay in $V-V_v$. This proves the inductive statement, and proves that there is no agency with $k\ge n-stretch_{YY}(G)$. 

To prove the equality in Theorem \ref{cl15}, we need to consider two cases, similar to those consider in the proof of Theorem \ref{cl9}. We define $u,v, i, j, H_u, H_v$ just the same as in that proof. We have two cases: either A) there are at least two distinct components say $H_1,H_2$ that contain at least one unoccupied node, or C) all unoccupied nodes are in the same component, say $H_1$. Case A) can be handled just the same as in the proof of Theorem \ref{cl9}. 

Case C) is very similar to Case B) of Theorem \ref{cl9}, but we will elaborate on that a bit. The main case is when there is at least one node of degree at least 3 in $H_1$, and assume that the nearest to $u$ is $u_1$. The distance between $u$ and $u_1$ is spanned by a path $P$ of at most $stretch_{YY}(G)$ nodes. Thus there are at least $stretch_{YY}(G)+1$ unoccupied nodes in $H_1$, and we may assume that they are arranged in the way of Figure \ref{CaseC}. That is, all nodes of $P-u$ are unoccupied, and $u_1$ has two more unoccupied neighbors, $u_{11},_{12}$. Please refer to Figure \ref{CaseC} for the way how to make moves (similar to Case B) to exchange $u$ and $v$, and thus prove that any agent can get anywhere. By applying this procedure repeatedly, we can construct an agency of $k$ agents, where $k$ is determined by the above formula.

\section*{Acknowledgments}
The authors are grateful for discussions on the topic with Tam\'as Kir\'aly and Zolt\'an Kir\'aly. The research was supported by the MTA-ELTE Egerv\'ary
Research Group and the Hungarian National Research, Development and Innovation Office NKFIH grant K109240.

\bibliographystyle{amsplain} 
\bibliography{stsp}

\end{document}